\documentclass[%
 reprint, superscriptaddress,
 amsmath,amssymb,
 aps,
 onecolumn,
]{revtex4-2}

\usepackage{graphicx}
\usepackage{dcolumn}
\usepackage{bm}
\usepackage{hyperref}
\usepackage{subfig}
\usepackage{color}
\usepackage{mathtools}

\begin{document}

\preprint{APS/123-QED}

\title{Confined active matter in external fields}

\author{Vaseem A. Shaik}
\affiliation{Department of Mechanical Engineering, Institute of Applied Mathematics,\\
University of British Columbia, Vancouver, BC, V6T 1Z4, Canada}
\author{Zhiwei Peng}
\affiliation{Division of Chemistry and Chemical Engineering, California Institute of Technology,\\
Pasadena, California 91125, USA}
\author{John F. Brady}
\affiliation{Division of Chemistry and Chemical Engineering, California Institute of Technology,\\
Pasadena, California 91125, USA}
\author{Gwynn J. Elfring}%
 \email{gelfring@mech.ubc.ca}
\affiliation{Department of Mechanical Engineering, Institute of Applied Mathematics,\\
University of British Columbia, Vancouver, BC, V6T 1Z4, Canada}

\date{\today}

\begin{abstract}
We analyze a dilute suspension of active particles confined between walls and subjected to fields that can modulate particle speed as well as orientation. Generally, the particle distribution is different in the bulk compared to near the walls. In the bulk, particles tend to accumulate in the regions of low speed, but in the presence of an orienting field, particles rotate to align with the field and accumulate downstream in the field direction. At the walls, particles tend to accumulate pointing into the walls and thereby exert pressure on walls. But the presence of strong orienting fields can cause the particles to reorient away from the walls, and hence shows a possible mechanism for preventing contamination of surfaces. The pressure at the walls depends on the wall separation and the field strengths. This work demonstrates how multiple fields with different functionalities can be used to control active matter under confinement.
\end{abstract}

\maketitle

\section{Introduction}
Active matter refers to a suspension of active particles that convert stored energy to directed motion \citep{Schweitzer2007}. Examples include a school of fish, a flock of birds and a suspension of microorganisms. Our focus here is on active matter systems where the active particles are micron sized and hence the inertia of the particles and the induced flow is negligible. Such active matter systems exhibit rich phenomena due to the self-propelling constituents, including collective motion \citep{Toner2005}, active turbulence \citep{Alert2022} and motility-induced phase separation \citep{Cates2015}.

The dynamics of active particles depend on characteristics such as their speed \citep{Volpe2011, Buttinoni2012, Palacci2013, Ross2019, Zhang2021}, orientation \citep{Takatori2014} and diffusivity \citep{Fernandez-Rodriguez2020}, and thus a degree of control over active matter can be exerted through modulation of these properties. By subjecting active particles to external fields like magnetic fields or gravitational fields or even gradients in heat, light or fluid viscosity, active particles have been shown to perform \textit{taxis} either by rotating to align with the external field or speeding up (or slowing down) in the field, or both. An example is \textit{Chlamydomonas nivalis} which reorients to preferentially swim against gravity due to bottomheaviness \citep{Kessler1985}. More impressively, one can `paint' with the bacterium \textit{E. coli} by exposing the bacterial suspension to the light gradients \citep{Arlt2018} as the bacterium changes its speed in response to light. More sophisticated control of synthetic active matter has recently been demonstrated by employing external magnetic fields and discrete-time feedback loops to tune the rotational diffusivity of active colloids \citep{Fernandez-Rodriguez2020}. It is in this vein, namely controlling the dynamics of active matter, that we develop theory for confined active particles subject to fields that can modulate particle speed as well as orientation.

Several researchers have analyzed the dynamics of active particles with spatially varying speeds \citep{Schnitzer1993, Tailleur2008, Cates2013, Cates2015, Arlt2019, Row2020}. When the spatial variation is slow and restricted to 1D, the number density $n$ and the swim speed $U$ are shown to be inversely proportional, $nU = constant$ \citep{Schnitzer1993}. This relation means that particles accumulate in the regions of low speed and has been shown to apply to active matter under abrupt speed changes provided that (thermal or biological) fluctuations are relatively weak \citep{Row2020}.

Active particles also tend to accumulate at confining boundaries due to their directional persistence \citep{Yan2015a}, and exert a force (or pressure) on these confining boundaries. This pressure, in the absence of any external field, is a sum of the bulk osmotic pressure and the swim pressure \citep{Yan2015}, which is the unique pressure required to confine active particles \citep{Takatori2014a}. If an orienting field is also present then the particles rotate to align with the field, net polar order develops and boundary accumulation is modified \citep{Yan2015a}.  Net polar order gives rise to a net average swim force and the wall pressure in this case is a sum of the swim pressure and the effective body force due to reorientation that acts on particles \citep{Yan2015a}. It has also been shown that in a bipolar orienting field, particles rotate to align along as well as against the field. This results in zero polar order but net nematic order. In such fields, the swim stress was shown to be tensorial and hence, the wall pressure is the wall normal component of the swim stress \citep{Yan2018}.

While particle speed and orientation can both be controlled by external fields \citep{Takatori2014a}, theory has not yet been developed for confined active matter subjected to external fields that modulate both. Specifically, the scaling law satisfied by the number density is not known, and thus previous experimental work compared results with the $nU = constant$ scaling law \citep{Stehnach2021}. The theory developed in previous work for the confined active matter in the presence of a orienting field was valid for wall separations much larger the run length, which is the distance an active particle travels before reorienting due to the rotary Brownian motion \citep{Yan2015a}. Here we derive theory for active matter subjected to the two aforementioned fields, valid in the relevant limit of weak translational Brownian motion (or  high activity \citep{Takatori2016}) and for all wall separations relative to the run length. We solve for the number density and wall pressure, as well as probe the theory and underlying physics that impact wall accumulation.

\begin{figure}[t!]
\centering
\includegraphics[scale = 0.57]{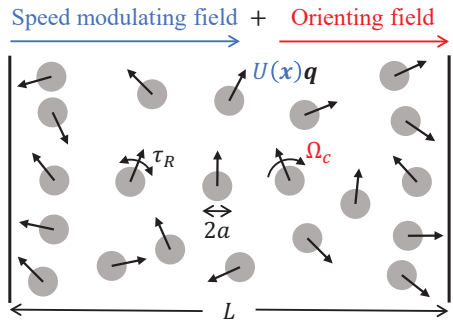}
\caption{A dilute suspension of spherical active particles of radius $a$ confined between two walls that are separated by a distance $L$. The particles are subjected to a field that modulates their speed spatially ${\bf{U}} = U\left( {\bf{x}} \right){\bf{q}}$ and also to a field that rotates them at a rate $\left|{\Omega _c} \right|$. Additionally, the particles also rotate at a rate $1/\tau_R = D_R$ due to an internal reorientation mechanism.}
\label{fig:fig1}
\end{figure}

\section{Confined active particles}
\subsection{Active Brownian particles}
We consider a dilute suspension of active particles confined between two infinite plane walls that are separated by a distance $L$. See Fig.~\ref{fig:fig1} for the schematic. The particles are subjected to an external field that modulates the particles' speed spatially ${\bf{U}} = U\left( {\bf{x}} \right){\bf{q}}$, where the unit vector ${\bf{q}}$ is the particle orientation. This can be achieved by imposing light on photo-sensitive bacteria or synthetic robots or even by spatially varying the `fuel' that bacteria consume. The particles are also subject to an external field that leads to reorientation to align with the field direction ${\bf{\hat H}}$. This reorientation can be caused by imposing magnetic or gravitational fields on the magnetotactic or bottom heavy bacteria, respectively, or even by spatially varying the background fluid viscosity \citep{Datt2019, Shaik2021}. The rate of reorientation is quantified by a characteristic angular velocity $\bm{\Omega}  = {\Omega _c}( {{\bf{q}} \times {\bf{\hat H}}})$, that is determined from the balance between torque caused by the field (which may be hydrodynamic or external depending on the particular mechanism driving reorientation) and rotational drag. For simplicity and with no lack of generality we assume the characteristic rate ${{\Omega _c} > 0}$.

The particles are subject to fluctuations that lead to translational and rotational diffusion with diffusivities, ${D_T}$ and ${D_R}$, respectively. The fluctuations may be thermal or biological in origin but regardless of the origin we consider the diffusivities to be constant and independent of the imposed background fields. Importantly, the particles in our model do not interact with one anther, hydrodynamically or otherwise. This simple model of active particles is called the active Brownian particle (ABP) model and it has been used widely to understand various phenomena without any hydrodynamic interactions \citep{bechinger16}.

\subsection{Kinetic theory}
We use a kinetic theory approach developed by \citet{Saintillan2013} to describe confined active matter in external fields. In this approach, the probability of finding a particle in the vicinity of position ${\bf{x}}$, and orientation ${\bf{q}}$ at time $t$, $P\left( {{\bf{x}},{\bf{q}},t} \right)d{\bf{x}}d{\bf{q}}$, is governed by the Smoluchowski equation
\begin{equation}
\frac{{\partial P}}{{\partial t}} + \nabla  \cdot {{\bf{j}}_T} + {\nabla _R} \cdot {{\bf{j}}_R} = 0,
\label{eqn:Smoluchowski}
\end{equation}
where ${\nabla _R} = {\bf{q}} \times \partial /\partial {\bf{q}}$ and the translational and rotational fluxes are ${{\bf{j}}_T} = {\bf{U}}\,P - {D_T}\nabla P$ and ${{\bf{j}}_R} = {\bm{\Omega }}\,P - {D_R}{\nabla _R}P$, respectively. The particles are prevented from entering the walls by enforcing zero translational flux normal to the wall ${\left. {{\bf{n}} \cdot {{\bf{j}}_T}} \right|_{{\rm{wall}}}} = 0$, where $\bf{n}$ is the unit vector normal to the wall. Also, the total number of particles is conserved by requiring that $\int {\int {Pd{\bf{q}}d{\bf{x}}} }  = 1$.

To capture the essential physics, we focus on the orientational moments of the probability density $P\left({\bf{x}}, {\bf{q}}, t\right)$. The first few moments are the number density $n = \int {Pd{\bf{q}}} $, the polar order ${\bf{m}} = \int {P{\bf{q}}d{\bf{q}}} $, and the nematic order ${\bf{Q}} = \int {P\left( {{\bf{qq}} - \frac{{{\bf{I}}}}{d}} \right)d{\bf{q}}} $, where $d$ is the dimensionality of the problem. These moments emerge naturally expanding the probability density in terms of the irreducible tensors of the orientation $\bf{q}$, $P\left({\bf{x}}, {\bf{q}}, t\right) = n + {\bf{m}} \cdot {\bf{q}} + {\bf{Q}}: \overbracket{{\bf{qq}}} + O\left(\overbracket{{\bf{qqq}}}\right)$, where the overbracket denotes the irreducible part of a tensor \citep{hess15}. Equations governing these moments can be derived by projecting \eqref{eqn:Smoluchowski} onto the basis of these irreducible tensors. Hence, $n$ and ${\bf{m}}$ satisfy
\begin{gather}
\frac{{\partial n}}{{\partial t}} + \nabla  \cdot {{\bf{j}}_n} = 0,
\label{eqn:n-Transport}\\
\frac{{\partial {\bf{m}}}}{{\partial t}} + \nabla  \cdot {{\bf{j}}_m} + {\Omega _c}{\bf{\hat H}} \cdot {\bf{Q}}
- {\Omega _c}n{\bf{\hat H}}\left( {1 - \frac{1}{d}} \right) + {D_R}\left( {d - 1} \right){\bf{m}} = {\bf{0}},
\label{eqn:m-Transport}
\end{gather}
where,
\begin{gather}
{{\bf{j}}_n} = \int {{\bf{j}}_T\,d{\bf{q}}}  = U\left( {\bf{x}} \right){\bf{m}} - {D_T}\nabla n,\\
{{\bf{j}}_m} = \int {{\bf{j}}_T \, {\bf{q}}d{\bf{q}}}  = U\left( {\bf{x}} \right){\bf{Q}} + \frac{{nU\left( {\bf{x}} \right)}}{d}{\bf{I}} - {D_T}\nabla {\bf{m}}.
\end{gather}
Similarly, moments of the no-flux condition at the walls ${\left. {{\bf{n}} \cdot {{\bf{j}}_T}} \right|_{{\rm{wall}}}} = 0$ yields ${\left. {{\bf{n}} \cdot {{\bf{j}}_n}} \right|_{{\rm{wall}}}} = 0$, ${\left. {{\bf{n}} \cdot {{\bf{j}}_m}} \right|_{{\rm{wall}}}} = 0$, for the first two moments. Finally, the integral constraint $\int {\int {Pd{\bf{q}}d{\bf{x}}} }  = 1$ implies $\int {n\,d{\bf{x}}}  = 1$.

\subsection{Analysis}
We begin first by examining the relevant physical scales in the problem. There are two time scales: the reorientation time, ${\tau _R} = 1/{D_R}$, due to rotary Brownian motion or some internal biological mechanism, and the time that the field takes to reorient the particle $1/ \Omega _c$. There are also three length scales: the microscopic length $h  = \sqrt {{D_T}{\tau _R}} $, the run length $\ell = {U_0}{\tau _R}$, and the channel width $L$, where $U_0$ is the self-propulsion speed in the absence of fields. We use the intrinsic reorientation time $\tau_R$, the run length $\ell$, and the speed in the absence of fields $U_0$ to non-dimensionalize the variables. The governing equations are ultimately characterized by three non-dimensional numbers: the P\'eclet number ${\rm{Pe}} = U_0\ell/{D_T}$ measuring the ratio of the self-advective (i.e., swimming) to the diffusive transport rate of particles, and two dimensionless groups which give the relative magnitude of the effects of the external fields: the relative importance of variations in speed ${\alpha _L} = \Delta U/{U_0}$, where $\Delta U$ is the characterstic change in speed, and the relative importance of the orienting field $\chi_R = \Omega_c \tau_R$. 

We simplify the analysis by focusing only on the steady state solutions and consider fields that are normal to the walls. We also consider only linear speed variations; hence, for a wall normal direction ${{\bf{e}}_x}$ (or $- {{\bf{e}}_x}$), the speed varies only along $x$, ${\bf{U}} = U\left( x \right){\bf{q}}$, where $U\left( x \right) = 1 - {\alpha _L}\left( {\frac{x}{L} - \frac{1}{2}} \right)$, and the particle rotates to align with ${\bf{\hat H}} = {{\bf{e}}_x}$. Such wall normal aligned fields mean that there is no physical mechanism to induce polar order parallel to the wall or to cause any variation in that direction. Hence $n = n\left( x \right)$, ${\bf{m}} = {m_x}\left( x \right){{\bf{e}}_x}$, and ${\bf{Q}} = {Q_{xx}}\left( x \right){{\bf{e}}_x}{{\bf{e}}_x} - \frac{{{Q_{xx}}\left( x \right)}}{{\left( {d - 1} \right)}}\left( {{\bf{I}} - {{\bf{e}}_x}{{\bf{e}}_x}} \right)$. Nematic order must be non-zero along the wall as $\bf{Q}$ is trace-free.

In general, in this problem nematic order is small $\left(\ll n\right)$ everywhere except possibly at the walls and there it remains small provided that P\'eclet numbers are modest $ {\rm{Pe}} < 10^3$ and the effects of the external fields are not dominant $\alpha_L < 1, \chi_R < 1$. See Fig.~\ref{fig:fig2_sm} in Appendix \ref{app:nematic} where we plot the nematic order $Q_{xx}$ as a function of position for various values of Pe, $\alpha_L$, and $\chi_R$, obtained from the full numerical solution of the Smoluchowski equation. Focusing (unless otherwise specified) on this range of parameter values, we neglect the nematic order, assuming $Q_{xx}=0$, to develop an analytical theory.

Active matter systems tend to have reasonably high P\'eclet numbers \citep{Takatori2016}, in which case the dominant transport process depends on the vicinity from the walls. In the bulk, away from the walls, advection is dominant, but near the walls, both advection and diffusion are equally important. To capture this, we perform a singular perturbation in $\rm{Pe}^{-1}$ and solve \eqref{eqn:n-Transport}, and \eqref{eqn:m-Transport} separately in the bulk and in the near wall boundary layer (BL) regions, with an appropriate matching of the resulting solutions. This perturbative analysis is valid provided the BL thickness $\lambda ^{-1}$ is small relative to the channel width $L$, $\lambda^{-1} \ll L$, in other words ${\rm{Pe}}L \gg 1$ because $\lambda \sim {\rm{Pe}}$.

\section{Results}
Regardless of Pe, there cannot be any particle flux normal to the wall, not just at the wall but anywhere in the domain, ${\bf{n}}\cdot {\bf{j}}_n = j_{n,x} = 0$ $\forall\, {\bf{x}}$. To derive this, integrate \eqref{eqn:n-Transport}, $\nabla \cdot {\bf{j}}_n = \frac{d}{dx}j_{n,x} = 0$ using the constraint ${\left. {{\bf{n}} \cdot {{\bf{j}}_n}} \right|_{{\rm{wall}}}} = \left. j_{n,x} \right|_{\rm{wall}} = 0$.

In the bulk, neglecting translational diffusion in the flux ${\bf{j}}_n = j_{n,x} {\bf{e}}_x = {\bf{0}}$, we get $U\left(x\right) m_x = 0$. Assuming the self-propulsion speed is never zero, we find that there is no polar order in the bulk, $m_x = 0$. This is unlike the situation in the absence of walls, where net polar order exists in the presence of an orienting field \citep{Takatori2014}, causing a finite particle flux. As there cannot be any particle flux in the presence of walls at steady state, there cannot be any polar order in the bulk either. Similarly, neglecting polar order and diffusion in the bulk in \eqref{eqn:m-Transport} we obtain
\begin{equation}
\frac{d}{{dx}}\left( {nU\left( x \right)} \right) - {\chi _R}n\left( {d - 1} \right) = 0.
\label{eqn:n-Bulk}
\end{equation}
The solution of this equation furnishes the number density in the bulk.

Unlike in the bulk, there tends to be polar order at the walls. Particles accumulate at the walls due to their persistent motion and on average, they are aligned into the walls simply because those aligned out of the walls swim away. However, the addition of strong enough orienting field can be used to rotate the particles away from the walls, ultimately preventing any wall accumulation. This points to a possible mechanism to prevent the contamination of surfaces.

To examine the boundary layer at the left wall, we rescale the position with the BL thickness, $\bar x = {\lambda _l}x$. We eliminate $m_x$ from ${{\bf{j}}_n} = {\bf{0}}$ and \eqref{eqn:m-Transport}, and evaluate the speed at the left wall $U_l$ to obtain
\begin{equation}
\left( {\frac{{{U_l}}}{d} + \frac{{\left( {d - 1} \right)}}{{{\rm{Pe}}{U_l}}}} \right)\frac{{dn}}{{d\bar x}} - \frac{{\lambda _l^2}}{{{\rm{P}}{{\rm{e}}^2}{U_l}}}\frac{{{d^3}n}}{{d{{\bar x}^3}}} - \frac{{{\chi _R}}}{{{\lambda _l}}}n\left( {1 - \frac{1}{d}} \right) = 0.
\label{eqn:n-BLLeft}
\end{equation}
Balancing advection with diffusion in this equation gives the BL thickness at the left wall $\lambda _l^2\sim {\rm{P}}{{\rm{e}}^2}\left( {\frac{{U_l^2}}{d} + \frac{{\left( {d - 1} \right)}}{{{\rm{Pe}}}}} \right)$. The BL thickness is the same at both walls in the absence of the fields, ${\lambda ^{ - 1}} = \lambda _l^{ - 1} = \lambda _r^{ - 1}$. On the other hand, balancing the reorientation with the diffusion gives the strength of the orienting field that is required to prevent any accumulation at the left wall, ${\chi _R}\sim {\lambda _l}\frac{{\lambda _l^2}}{{{\rm{P}}{{\rm{e}}^2}{U_l}}}\sim {\rm{Pe}}$. Similar analysis can also be carried out in the BL at the right wall.

We solve the equations in the bulk and those in the BLs and form a composite expansion. We do this calculation in a number of limits, namely: no external fields $\left( {U = 1,{\chi _R} = 0} \right)$, a weak speed modulating field $\left( {{\chi _R} = 0,{\alpha _L} \ll 1} \right)$, and also a moderate orienting field $\left( {U = 1,\,\,{\chi _R} < 1} \right)$. This theory is not valid for strong orienting fields $\left( {U = 1,\,\,{\chi _R} \gg 1} \right)$ as nematic order becomes large and cannot be neglected; thus, we develop an alternative theory for strong orienting fields. We validate these theories by comparing them with 2D Brownian Dynamics (BD) simulations. See Appendices \ref{app:theory}, \ref{app:BD}, respectively, for the exact theoretical expressions and the BD simulation procedure.

\begin{figure}[tbh!]
\subfloat{\includegraphics[scale = 0.61]{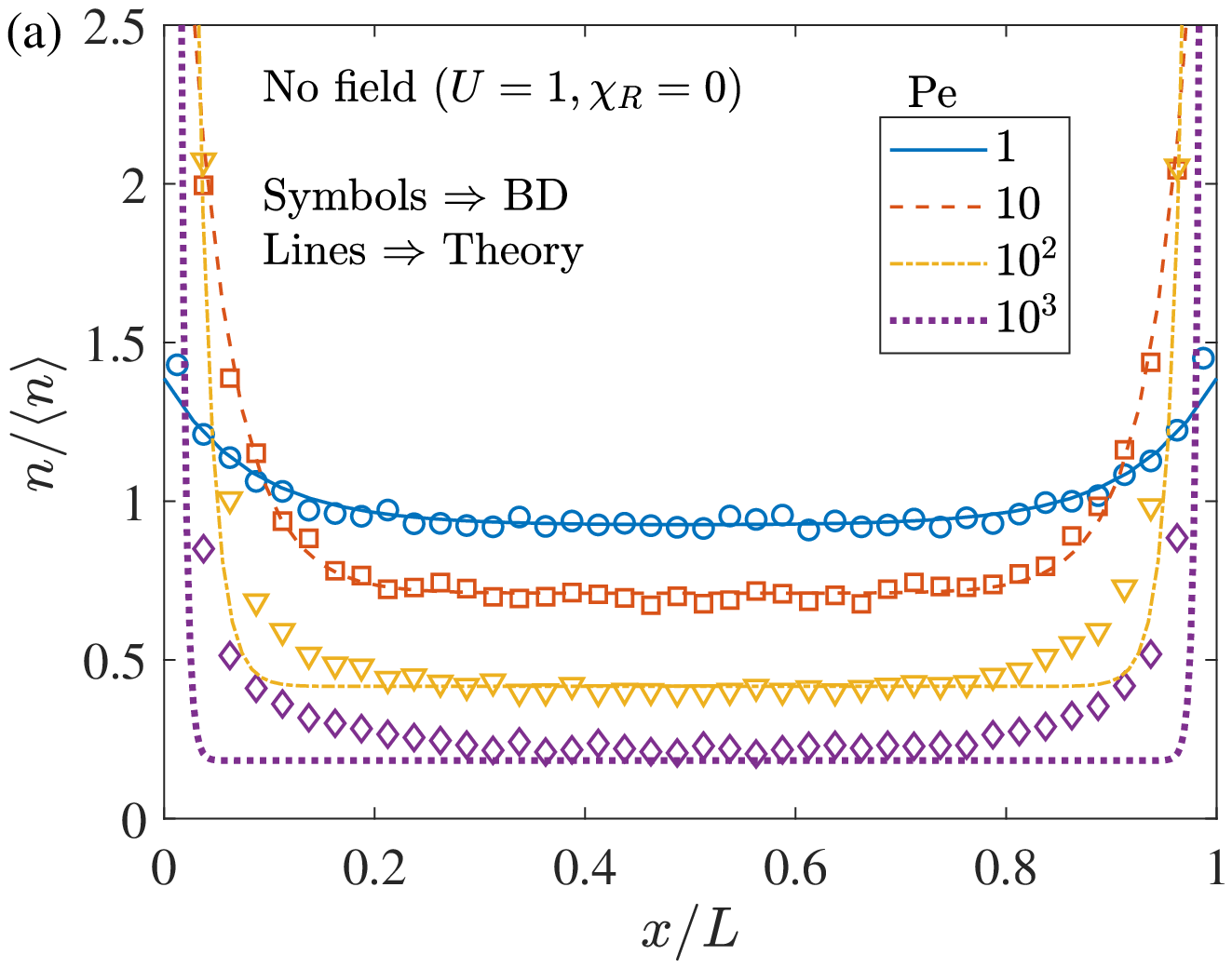}\label{fig:fig2a}}
\subfloat{\includegraphics[scale = 0.61]{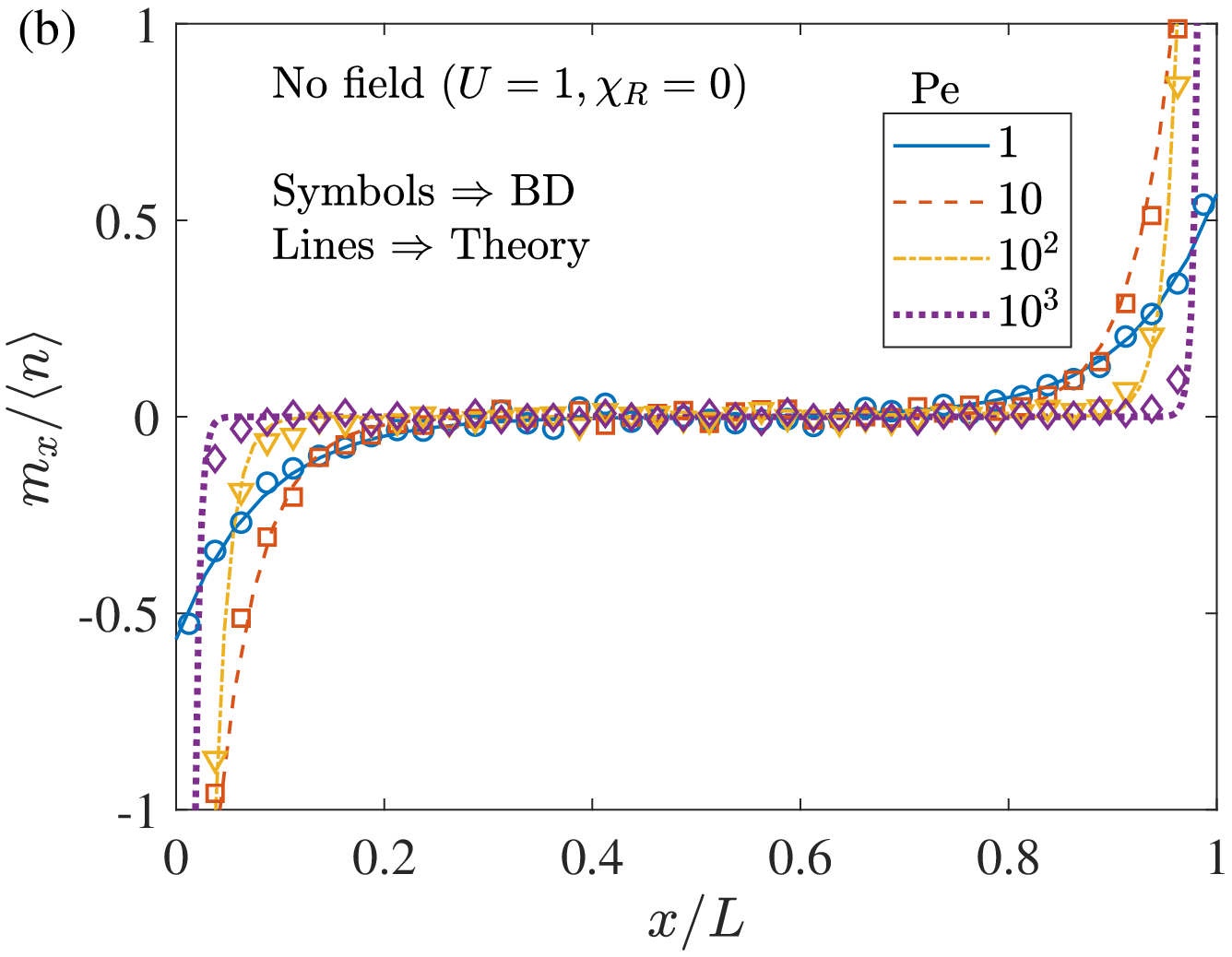}\label{fig:fig2b}}
\caption{\label{fig:fig2} In the absence of any field, the number density (a) and the polar order (b) associated with the active matter system. Here, the symbols denote the BD simulation results while the lines represent the theory (see \eqref{eqn:n-NofieldFinal}, \eqref{eqn:mx-Nofield} in Appendix \ref{app:theory}). Also, the confinement region $L$ is 10 times larger than the microscopic length $h$.}
\end{figure}

\subsection{No external fields}
In the absence of any external field, the self-propulsion speed $U = 1$ and the reorientation parameter $\chi_R = 0$. Then in the bulk, while the polar order is zero, the number density satisfies $nU = constant$ or simply $n = constant$ from \eqref{eqn:n-Bulk} (see also Fig.~\ref{fig:fig2}). On the other hand, at the walls, the particles accumulate and align into the walls, hence, $m_x < 0$ at the left wall and $m_x> 0$ at the right wall. Both the number density and polar order at the wall increase while the thickness of the boundary layer decreases with increasing Pe, $\left( {{\lambda ^{ - 1}}\sim {\rm{P}}{{\rm{e}}^{ - 1}}} \right)$, while the bulk concentration decreases to conserve the total number density. In this limit, the theory developed here is consistent with previous works \citep{Yan2015, Row2020}. It is also in agreement with Brownian dynamics simulations at moderate $\text{Pe}$; at very high $\rm{Pe}$ the theory breaks down due to the failure of the zero nematic order closure used.

When active particles collide with walls they exert a force or pressure ($\sim$ force/area) on them. In the absence of external fields, the pressures exerted on the left wall and the right wall are the same, ${\Pi ^{LW}} = {n^{LW}}{k_B}T = {\Pi ^{RW}} = {n^{RW}}{k_B}T$ (reported here in dimensional form), where $n^{LW}$, $n^{RW}$ are the number densities at the left and right walls, respectively. This wall pressure is the sum of the osmotic pressure in the bulk and the swim pressure
\begin{equation}
{\Pi ^W} = {\Pi ^{LW}} = {\Pi ^{RW}} = {n^{bulk}}\left( {k_B T + k_sT_s^0} \right).
\end{equation}
This formula simplifies in the high activity limit $\left( {{\rm{Pe}} \gg 1} \right)$ after relating the bulk concentration ${n^{bulk}}$ to the average concentration $\left\langle n \right\rangle  = \frac{1}{L}\int_0^L {n\,dx}$ as \citep{Yan2015}
\begin{equation}
{\Pi ^W} = \frac{{\left\langle n \right\rangle \left( {k_BT + k_sT_s^0} \right)}}{{1 + \frac{2}{{L\sqrt d \left( {d - 1} \right)}}}}.
\end{equation}
Here, the particle activity is defined as ${k_s}{T_s^0} = {\rm{Pe}}{k_B}T/d\left( {d - 1} \right) = \zeta U_0^2 \tau_R/d\left( {d - 1} \right)$, where $\zeta$ is the drag coefficient, $k_B$ is the Boltzmann constant and $T$ is the absolute temperature.

\begin{figure*}[tbh!]
\centering
\subfloat{\includegraphics[scale = 0.43]{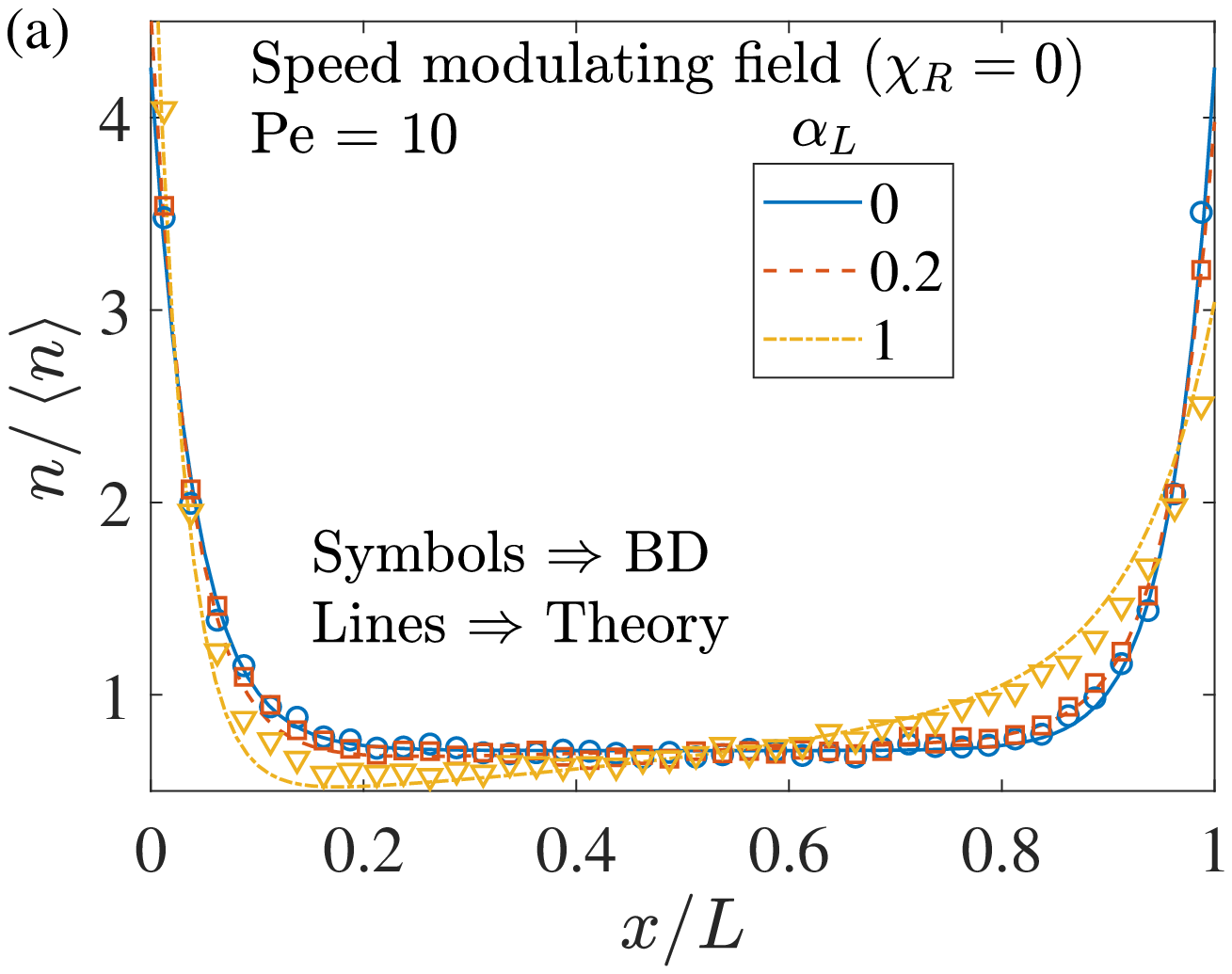}\label{fig:fig3a}}
\subfloat{\includegraphics[scale = 0.43]{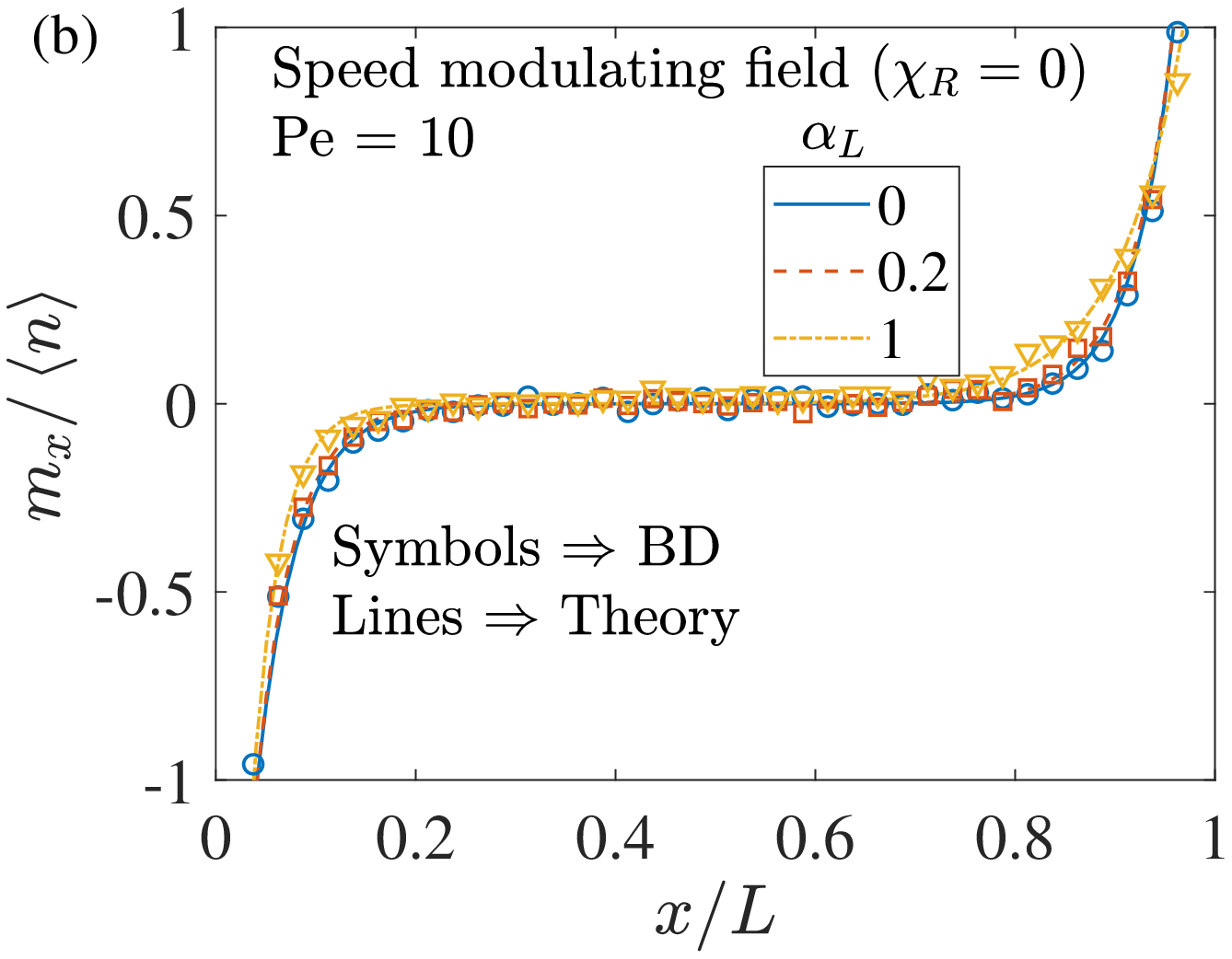}\label{fig:fig3b}}
\subfloat{\includegraphics[scale = 0.43]{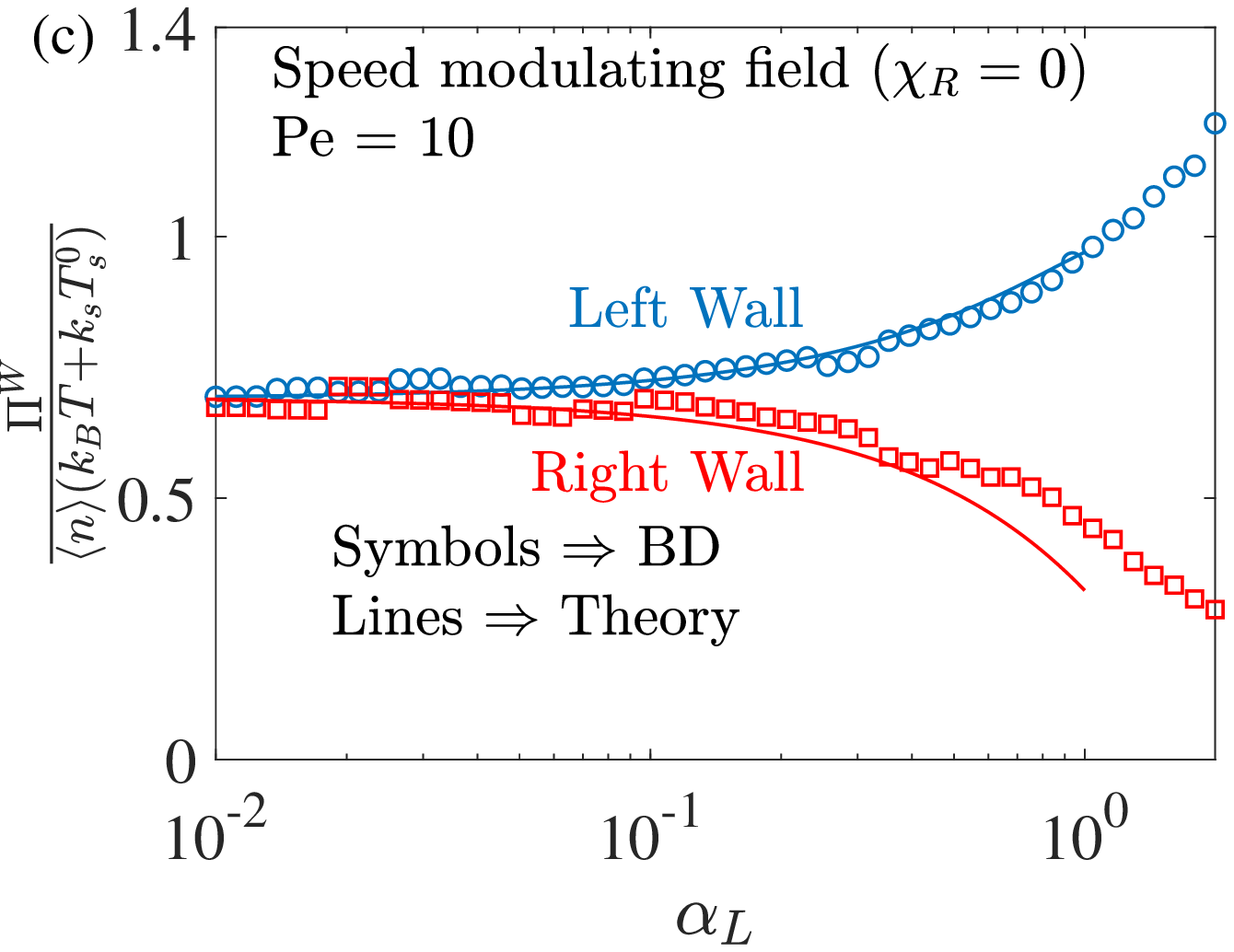}\label{fig:fig3c}}
\caption{\label{fig:fig3} The number density (a), the polar order (b), and the wall pressure (c) associated with the active matter subjected to the speed modulating field. The symbols denote the BD simulation results while the lines represent the theory. The confinement region $L$ is 10 times larger than the microscopic length $h$.}
\end{figure*}

\subsection{Spatially varying speed}
In the presence of a field that modulates the speed spatially, say $U = 1 - {\alpha _L}\left( {\frac{x}{L} - \frac{1}{2}} \right)$ but with $\chi_R=0$, the bulk polar order is still zero while the number density follows $nU = constant$ scaling from \eqref{eqn:n-Bulk}. This means that the particles in the bulk accumulate in the regions of low speed (see Fig.~\ref{fig:fig3a}). Additionally, the particles also accumulate at the walls. However, unlike in the bulk where the concentration decreases with increasing speed, the number density at the walls increases with higher particle speeds. Essentially this is because faster particles travel more quickly to the wall in comparison to slower ones. 

The speed and hence, the accumulation at the left wall increases (and those at the right wall decrease) with an increase in the field strength $\alpha_L$. These accumulated particles exert a pressure on the walls, and hence the pressure on the left and right walls correspondingly increase and decrease with increasing field strength (see Fig.~\ref{fig:fig3c}). In weak fields $\left( {{\alpha _L} \ll 1} \right)$, the wall pressure is again the sum of the bulk osmotic pressure and the swim pressure, but evaluated at the wall
\begin{equation}
\begin{split}
{\Pi ^{LW}} = {n^{bulk,LW}}\left( {k_B T + k_s T_s^{LW}} \right),\\
{\Pi ^{RW}} = {n^{bulk,RW}}\left( {k_B T + k_s T_s^{RW}} \right),
\end{split}
\end{equation}
and it simplifies at high activities $\left( {\rm{Pe}} \gg 1\right)$ to
\begin{equation}
\frac{{{\Pi ^{LW}}}}{{{U_l}}} = \frac{{{\Pi ^{RW}}}}{{{U_r}}} = \frac{{\left\langle n \right\rangle \left( {k_B T + k_s T_s^0} \right)}}{{\frac{1}{{{\alpha _L}}}\ln \left( {\frac{{{U_l}}}{{{U_r}}}} \right) + \frac{2}{{L\sqrt d \left( {d - 1} \right)}}}}.
\label{eqn:Wall-Pressure-Speed}
\end{equation}
Here, the activity ${k_s}{T_s}$ is defined locally (in dimensional form) as ${k_s}{T_s} = \zeta U{\left( x \right)^2}{\tau _R}/d\left( {d - 1} \right)$ and it simplifies in the absence of field to ${k_s}T_s^0 = \zeta U_0^2{\tau _R}/d\left( {d - 1} \right)$. The pressure imbalance $\left( \Pi^{LW} \ne \Pi^{RW} \right)$ and the resultant net force from the walls is balanced by the net swim force (that acts as a body force \cite{Yan2015a}), which can arise due to spatial variations in speed \cite{Takatori2015}, or orientation bias \cite{Yan2015a} as discussed below.

\begin{figure*}[t!]
\centering
\subfloat{\includegraphics[scale = 0.44]{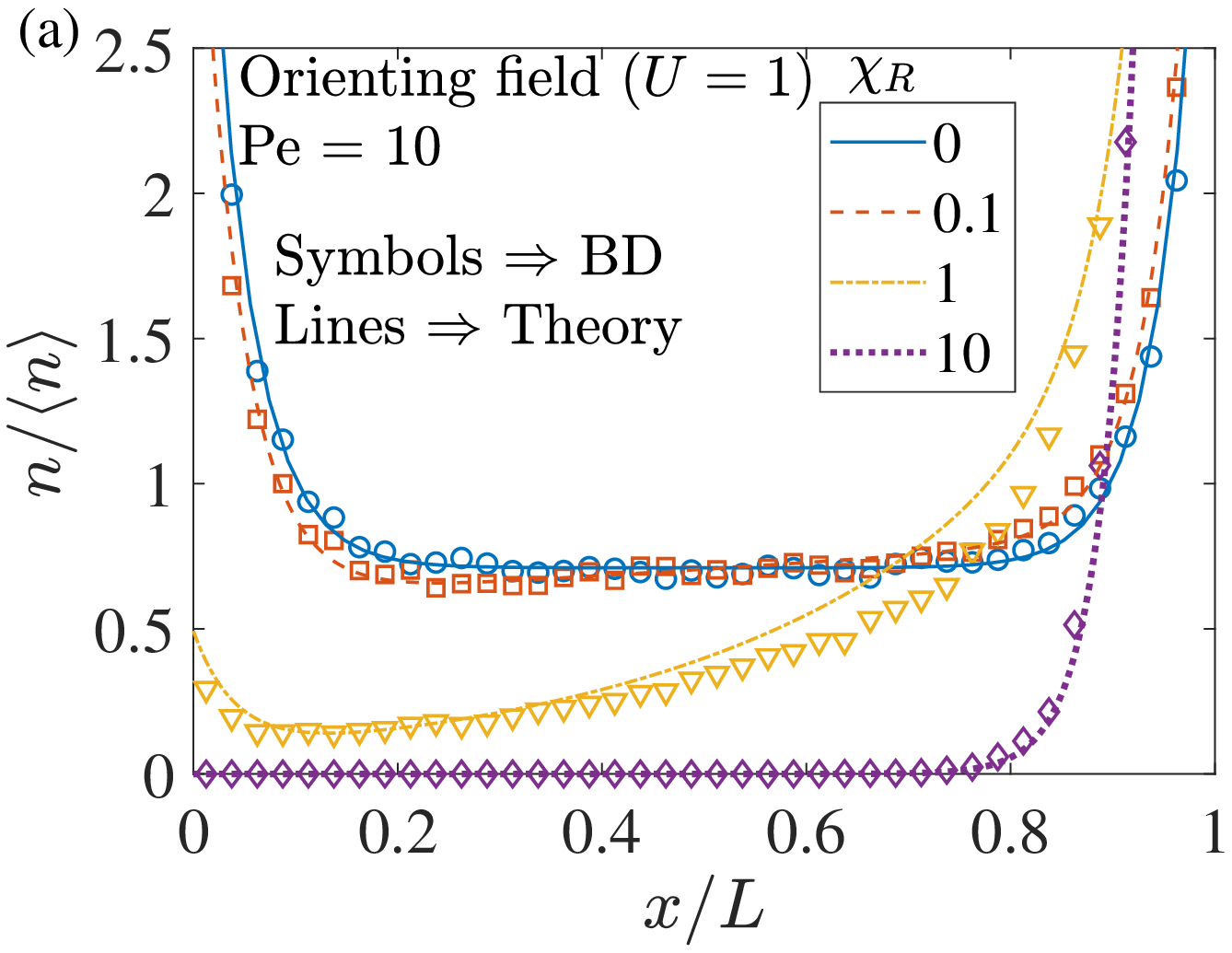}\label{fig:fig4a}}
\subfloat{\includegraphics[scale = 0.44]{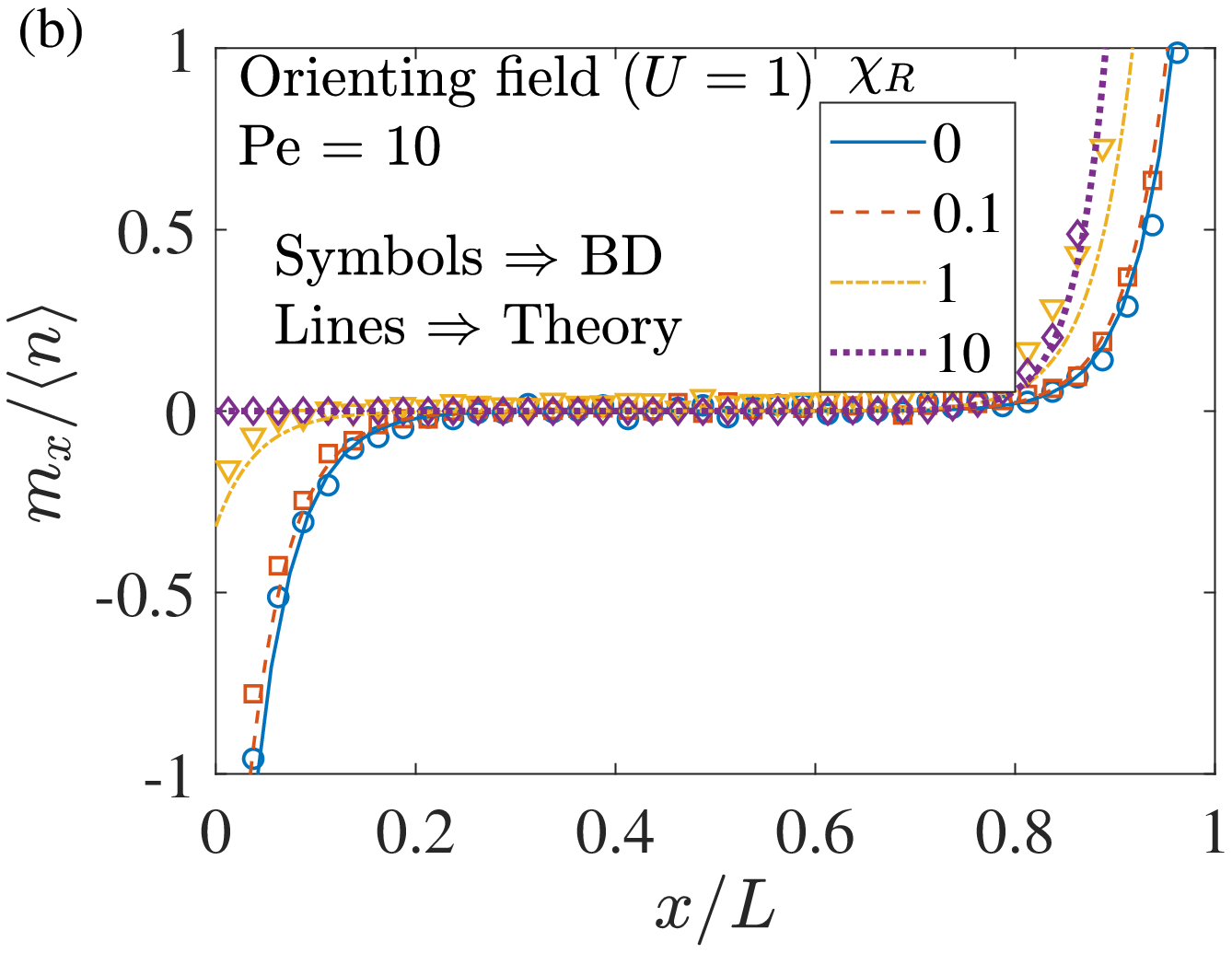}\label{fig:fig4b}}
\subfloat{\includegraphics[scale = 0.43]{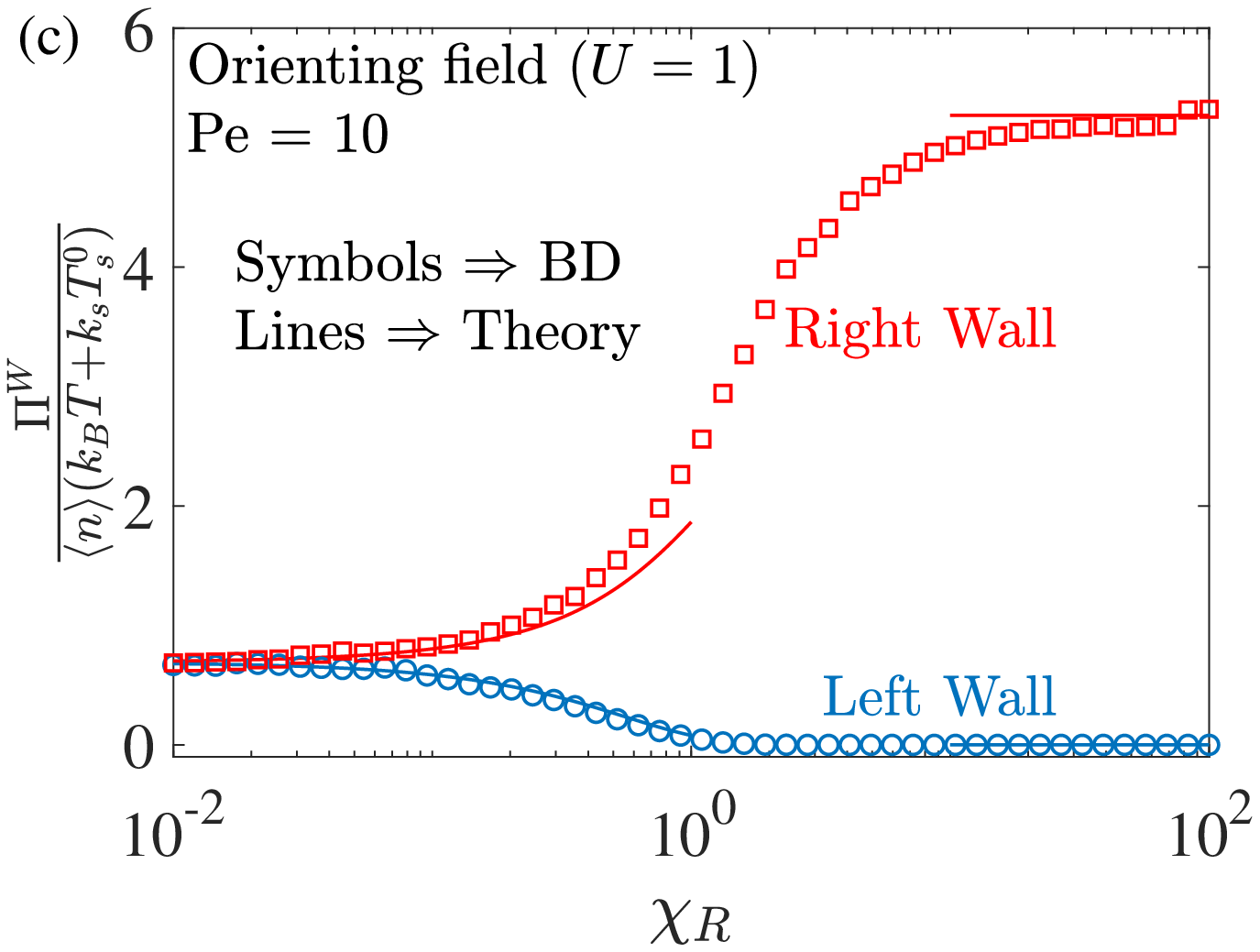}\label{fig:fig4c}}
\caption{\label{fig:fig4} The number density (a), the polar order (b), and the wall pressure (c) associated with the active matter subjected to the orienting field. The symbols denote the BD simulation results while the lines represent the theory. The confinement region $L$ is 10 times larger than the microscopic length $h$.}
\end{figure*}

\subsection{Particles in orienting fields}
In an orienting field $\left( {U = 1,{\chi _R} \ne 0} \right)$, the particles rotate to align with the field for $\chi_R > 0$. Then, while the bulk polar order has to be zero to enforce the zero particle flux at steady state, the number density follows the exponential distribution from \eqref{eqn:n-Bulk}, $n = constant \cdot {e^{\left( {d - 1} \right){\chi _R}x}}$. This means the particles in bulk accumulate downstream or at right for $\chi_R > 0$ (See Fig.~\ref{fig:fig4a}). When $\chi_R=0$ particles accumulate equally at both walls, but as $\chi_R$ increases, the accumulation at the right wall is increased while that at the left wall is diminished as particles are driven from left to right, the concentration and polar order thus become increasingly asymmetric as $\chi_R$ increases (see Fig.~\ref{fig:fig4a} and Fig.~\ref{fig:fig4b}).

The pressure exerted by the particles on the wall follows the same trend as the accumulation i.e., the pressure on the left and right walls, respectively, decrease and increase with increasing field strength $\chi_R$ (see Fig.~\ref{fig:fig4c}). Again, a simple expression for the wall pressure can be found in the limit of high activity $\left( {{\rm{Pe}} \gg 1} \right)$ and weak field $\left( {{\chi _R} \ll 1} \right)$
\begin{equation}
\frac{{{\Pi ^{LW}}}}{{{e^{ - \kappa }}}} = \frac{{{\Pi ^{RW}}}}{{{e^\kappa }}} = \frac{{\left\langle n \right\rangle \left( {k_B T + k_s {T_s^0}} \right)}}{{\frac{{\sinh \kappa }}{\kappa } + \frac{2}{{L\sqrt d \left( {d - 1} \right)}}\cosh \kappa }},
\label{eqn:Wall-Pressure-LowChiR}
\end{equation}
where $\kappa  = {\chi _R}\left( {d - 1} \right)L/2$.

\subsection{Strong orienting fields}
In strong orienting fields $\left( {{\chi _R} \gg 1} \right)$, the nematic order at the right wall becomes important and hence cannot be neglected. See Fig.~\ref{fig:fig2c_sm} in Appendix \ref{app:nematic}. As the theory developed here relies on the zero nematic order closure, it is not valid in this situation. However, some physical insights can still be drawn by applying the current theory in the strong field limit. In strong fields, for $\chi_R > 0$, we expect all the particles to align with the field, leave the left wall, and accumulate at the right wall. Hence, the left wall should be free of any particles while the accumulation at the right wall should asymptote to a value determined from the balance between the particle advection and diffusion there $\left( {{\rm{Pe}}\,n\sim \frac{{dn}}{{dx}}} \right)$. Similarly, the pressure acting on the left wall should be zero and that acting on the right wall should asymptote to a value that depends on the particle accumulation there. We next confirm these predictions by developing an alternative theory modeling the strong field limit.

Most particles in strong orienting fields $\left( {{\chi _R} \gg 1} \right)$ are aligned along the field. Hence, we approximate the probability density in this case as $P\left( {{\bf{x}},{\bf{q}}} \right) = n\left( {\bf{x}} \right)\delta \left( {{\bf{q}} - {\bf{\hat H}}} \right)$, where $\delta$ is the Dirac delta function \citep{saintillan08, gao17}. This reduces the polar and nematic order to ${m_x} = n$, ${Q_{xx}} = n\left( {1 - \frac{1}{d}} \right)$. Using these, we solve \eqref{eqn:n-Transport} by enforcing the constraints ${\left. {{\bf{n}} \cdot {{\bf{j}}_n}} \right|_{{\rm{wall}}}} = 0$ and $\frac{1}{L}\int {n\,d{\bf{x}}}  = \left< n\right>$, to ultimately find the number density that is correct at any ${\rm{Pe}} \le O\left( {{\chi _R}} \right)$
\begin{equation}
\frac{n}{\left<n\right>} = \frac{{{\rm{Pe}}L}}{{\left( {{e^{{\rm{Pe}}L}} - 1} \right)}}{e^{{\rm{Pe}}x}}.
\end{equation}
However, in the limit where the earlier singular perturbation analysis is valid, ${\rm{Pe}}L \gg 1$, the number density simplifies to
\begin{equation}
\frac{n}{\left< n\right>} = {\rm{Pe}}\,L\,{e^{ - {\rm{Pe}}\left( {L - x} \right)}}.
\label{eqn:HighChiRTheory}
\end{equation}
This equation predicts the particles are confined in a BL of thickness Pe$^{-1}$ at the right wall (see Fig.~\ref{fig:fig4a}). There are no particles left in the bulk or at the left wall, hence the pressure acting on the left wall is zero. On the other hand, the pressure exerted on the right wall can be found from \eqref{eqn:HighChiRTheory} as
\begin{equation}
{\Pi ^{LW}} = 0,\,\,{\Pi ^{RW}} = {\rm{Pe}}\,L \left<n\right>{k_B}T = d\left( {d - 1} \right)L \left<n\right>{k_s}{T_s^0}.
\label{eqn:Wall-Pressure-HighChiR}
\end{equation}

\subsection{Non-uniform orienting fields}

\begin{figure*}[tbh!]
\centering
\subfloat{\includegraphics[scale = 0.61]{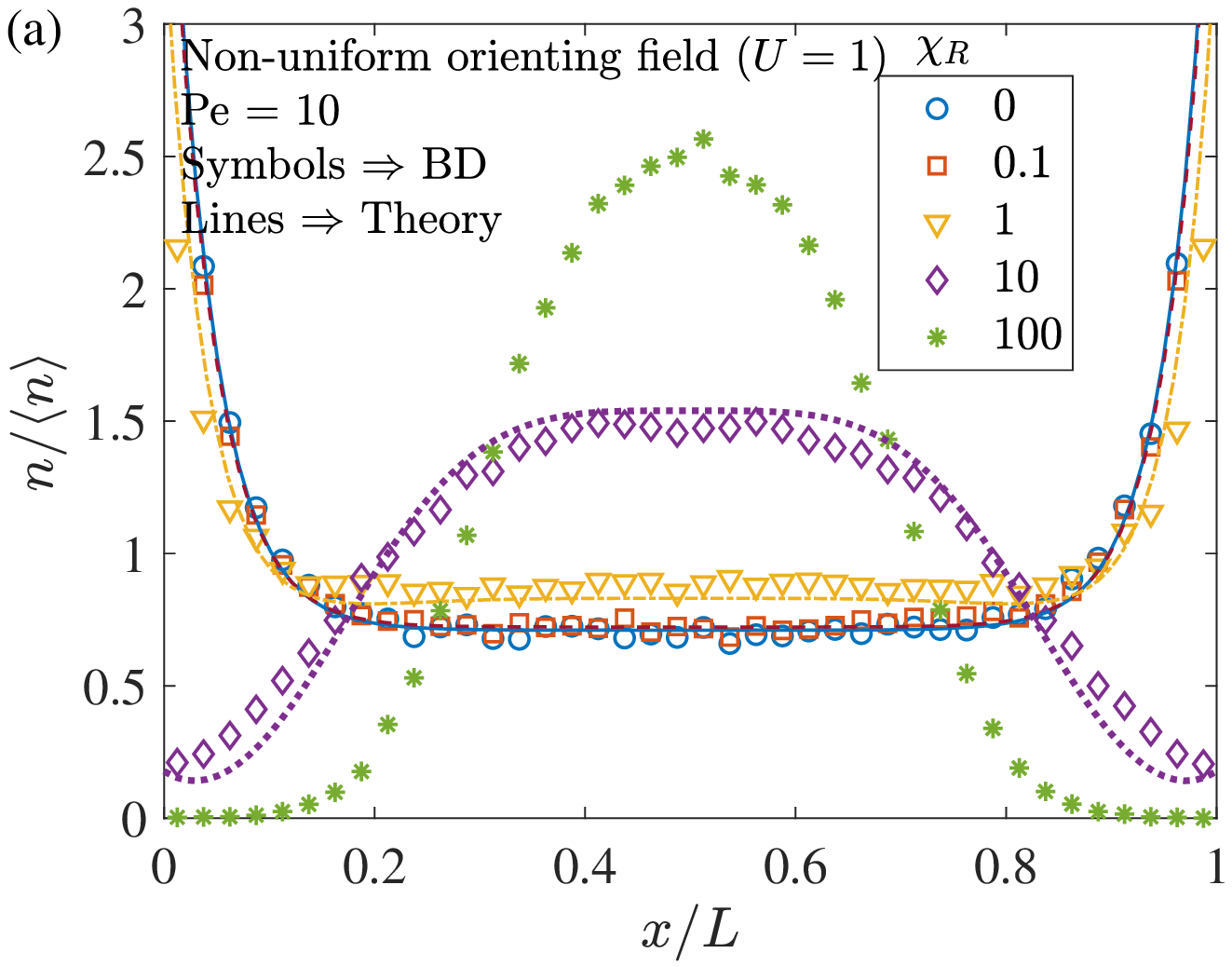}\label{fig:fig5a}}
\subfloat{\includegraphics[scale = 0.61]{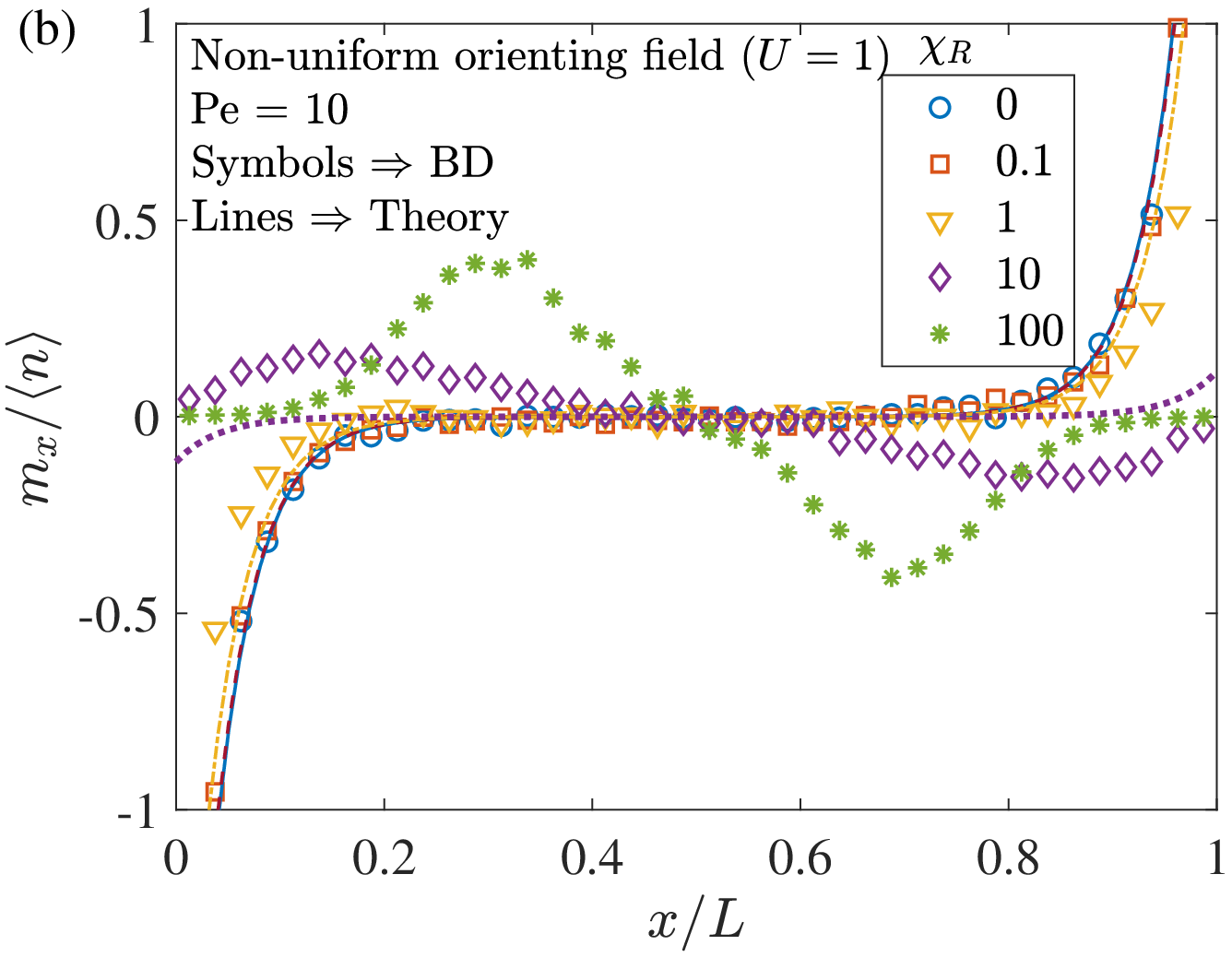}\label{fig:fig5b}}
\caption{\label{fig:fig5} The number density (a) and the polar order (b) associated with the active matter system subjected to non-uniform orienting field, ${\bf{H}} = - \left(2x/L - 1\right)^3 {\hat{\bf{H}}}$. The symbols denote the BD simulation results while the lines represent the theory. The confinement region $L$ is 10 times larger than the microscopic length $h$.}
\end{figure*}
In the uniform orienting fields $\bf{H} = \hat{\bf{H}}$, we discussed how the particles rotate to align with the field, ultimately leaving one wall and accumulating at the other wall in strong fields. This suggested the potential use of orienting fields in preventing accumulation at one of the walls. We can also prevent accumulation at both walls by using a non-uniform orienting field ${\bf{H}} = H\left( x\right) \hat{\bf{H}}$, where $H\left(x\right)$ is an odd function relative to the centerline $x=L/2$ i.e., $H\left(\frac{L}{2} - x\right) = - H\left(x - \frac{L}{2} \right)$. For $H\left(0\right) > 0$, this field points away from both walls, towards the center. In such fields, the particles rotate with velocity ${\bm{\Omega}} = \Omega_c H\left(x\right) \left({\bf{q}} \times \hat{\bf{H}}\right)$ to align with the field, ultimately leaving both walls and accumulating at the center if the field is strong enough. We demonstrate this behavior for a cubic function $H\left(x\right) = -\left(2x/L - 1 \right)^3$ in Fig.~\ref{fig:fig5}, where we see that the particles indeed move from the wall towards the center as the field strength $\chi_R$ increases. A moderate field theory $\left(\chi_R \ll {\rm{Pe}}\right)$ for non-uniform fields can also be developed by simply replacing the constant angular velocity $\Omega_c$ or $\chi_R$ in the formulae for uniform field (\eqref{eqn:m-Transport}, \eqref{eqn:n-Bulk}, \eqref{eqn:n-BLLeft})  with $\Omega_c H\left(x\right)$ or $\chi_R H\left(x\right)$. This still yields zero polar order in the bulk but a different number density $n = constant \cdot e^{\chi_R \left(d-1\right) \int{H\left(x\right) dx}}$.

\subsection{Combined effects}
With both speed modulating and uniform orienting fields, the physics is a combination (not necessarily a linear combination) of that for the individual fields. To illustrate this, we consider the behavior in the bulk. Here, with the speed modulating field, we know the particles accumulate in the regions of low speed. Hence, for a speed decreasing from left to right, the particles accumulate at right. On the other hand, in the orienting field, for $\chi_R < 0$, the particles rotate to align against the field and accumulate in the upstream of the field (left). Then in both fields, the particles either accumulate in the regions of low speed (right) or in the upstream of the orienting field (left) depending on the relative magnitude of the field strengths $\left( {{\chi _R}/{\alpha _L}} \right)$. Also, there can be no accumulation at all $\left( {n = constant} \right)$ if the opposing effects of the fields cancel each other. This discussion is indeed consistent with the exact expression for number density found from \eqref{eqn:n-Bulk}
\begin{equation}
n = \frac{{constant}}{{{U^{1 + \left( {d - 1} \right){\chi _R}L/{\alpha _L}}}}},
\end{equation}
where ${\chi _R}/{\alpha _L}$ has to be $ - 1/\left( {d - 1} \right)L$ in order for the fields to cancel each other. Also, as expected, this number density simplifies to $n = constant$ in the absence of both fields ${\left( {{\chi _R},{\alpha _L}} \right) \to \left( {0,0} \right)}$ and to $nU = constant$ or $n = constant \cdot {e^{\left( {d - 1} \right){\chi _R}x}}$ in the presence of the speed modulating field ${\chi _R} \to 0$ or the orienting field ${\alpha _L} \to 0$, respectively.

\section{Conclusions}
In summary, we analyzed confined active matter subjected to speed modulating and orienting fields. We showed that bulk polar order is always zero while the number density satisfied the usual $nU = constant$ scaling in the speed modulating fields but a different exponential distribution in orienting fields. The particles usually accumulate at the walls, but the orienting fields can be used to turn the particles away from the wall, ultimately preventing the accumulation at a wall. We also discussed the force exerted by the active matter on the confining walls and provided a concise expressions for the wall pressure.

Here we have neglected hydrodynamic interactions between the active particles and a natural next step is to include them. Hydrodynamic interaction between active particles are generally dipolar to leading order (in a dilute system) and hence determined by the $\mathbf{Q}$ tensor field \cite{Saintillan2013}, which therefore cannot be neglected and more sophisticated closures (besides simple truncation) are generally used to ameliorate the hierarchy problem associated with projection of the Smoluchoswki equation onto moments \cite{gao17,weady22}.


\section*{Acknowledgements}
Gwynn Elfring acknowledges the hospitality of the Division of Chemistry and Chemical Engineering at the California Institute of Technology during a sabbatical stay, supported by a UBC Killam Research Fellowship, that served as a formative period of this work and funding from the Natural Sciences and Engineering Research Council of Canada (RGPIN-2020-04850). 

\newpage

\appendix

\section{\label{app:theory}Analytical expressions for the number density and the polar order}
For active matter confined between walls at $x = 0$ and $L$, we find the number density and polar order by solving Eqs.~\eqref{eqn:n-Transport}, \eqref{eqn:m-Transport} in the main text along with the constraints ${\left. {{\bf{n}} \cdot {{\bf{j}}_n}} \right|_{{\rm{wall}}}} = 0$ and ${\left. {{\bf{n}} \cdot {{\bf{j}}_m}} \right|_{{\rm{wall}}}} = \bf{0}$. We solve these equations asymptotically by doing a singular perturbation expansion in Pe$^{-1}$ for ${\rm{Pe}}L \gg 1$. Additionally, in the absence of external field, an exact solution valid at any ${\rm{Pe}}L$ is also found. In either case, the aforementioned no-flux conditions determine solutions up to a multiplicative constant that is found using the additional constraint $\frac{1}{L}\int_0^L {n\,dx}  = \left\langle n \right\rangle$.

When there is no external field $\left( {U = 1,\,{\chi _R} = 0} \right)$, the exact solution is
\begin{equation}
\frac{n}{{{n_c}}} = \gamma b\left[ {\cosh \left( {\lambda \left( {x - \frac{L}{2}} \right)} \right) - 1} \right] + 1,
\label{eqn:n-Nofield}
\end{equation}
\begin{equation}
\frac{{{m_x}}}{{{n_c}}} = b\sinh \left( {\lambda \left( {x - \frac{L}{2}} \right)} \right),
\end{equation}
where
\begin{equation}
b = \frac{\gamma }{{d\left\{ {1 + \Lambda \left[ {\cosh \left( {\frac{{\lambda L}}{2}} \right) - 1} \right]} \right\}}},
\end{equation}
\begin{equation}
\lambda  = {\rm{Pe}}\sqrt {\frac{1}{d} + \frac{{\left( {d - 1} \right)}}{{{\rm{Pe}}}}} ,\,\,\Lambda  = \frac{1}{{1 + \frac{{{\rm{Pe}}}}{{d\left( {d - 1} \right)}}}},\,\,\gamma  = \sqrt {\frac{d}{{1 + \frac{{d\left( {d - 1} \right)}}{{{\rm{Pe}}}}}}} .
\label{eqn:Constants}
\end{equation}
Here, $n_c$ is the number density at the center of the confinement and can be found using the constraint $\frac{1}{L}\int_0^L {n\,dx}  = \left\langle n \right\rangle $. 
This exact solution is consistent with the previous calculation on confined active matter \citep{Row2020}. The number density in \eqref{eqn:n-Nofield} rewritten as 
\begin{equation}
\frac{n}{{n_0}} = 1 + \frac{{{\rm{Pe}}}}{d\left(d - 1\right)} \frac{\sinh\left(\lambda x\right) + \sinh\left(\lambda \left(L - x \right) \right)}{\sinh\left(\lambda L \right)}
\label{eqn:n-Nofield2}
\end{equation}
where $n_0 = \frac{n_c}{1 + \frac{{\rm{Pe}}}{d\left(d - 1\right)}\frac{1}{\cosh\left(\lambda L /2\right)}}$, is similar to that reported in Ref.~\citep{Yan2015}. 

The exact solution expanded in ${\rm{P}}{{\rm{e}}^{ - 1}}$ takes the form
\begin{equation}
\frac{n}{{{n^{bulk}}}} = 1 + \frac{{\rm{Pe}}}{{d\left( {d - 1} \right)}}\left\{ {{e^{ - \lambda x}} + {e^{ - \lambda \left( {L - x} \right)}}} \right\}  + O\left( {{\rm{P}}{{\rm{e}}^{ - 1}}} \right),
\label{eqn:n-NofieldFinal}
\end{equation}
\begin{equation}
\frac{{{m_x}}}{{{n^{bulk}}}} = \frac{\lambda }{{d\left( {d - 1} \right)}}\left\{ { - {e^{ - \lambda x}} + {e^{ - \lambda \left( {L - x} \right)}}} \right\}  + O\left( {{\rm{P}}{{\rm{e}}^{ - 1}}} \right)
\label{eqn:mx-Nofield}
\end{equation}
where the bulk number density ${n^{bulk}} = n_c$. Here, the leading order terms display a linear combination of the near-wall solution \citep{Yan2015} and the bulk solution.

In the presence of a field that modulates the self-propulsion speed (say $U = 1 - {\alpha _L}\left( {\frac{x}{L} - \frac{1}{2}} \right)$), the asymptotic solution for weak fields $\left( {{\alpha _L} \ll 1} \right)$ is
\begin{equation}
\frac{n}{{{n_c}}} = \frac{1}{U} + \frac{{\rm{Pe}}}{{d\left( {d - 1} \right)}}\left\{ {{U_l}{e^{ - {\lambda _l}x}} + {U_r}{e^{ - {\lambda _r}\left( {L - x} \right)}}} \right\} + O\left( {{\rm{P}}{{\rm{e}}^{ - 1}},\frac{{{\alpha _L}}}{{{\rm{Pe}}L}}} \right),
\end{equation}
\begin{equation}
\frac{{{m_x}}}{{{n_c}}} = \frac{1}{{d\left( {d - 1} \right)}}\left\{ { - \lambda_l {e^{ - \lambda_l x}} + {\lambda _r}{e^{ - {\lambda _r}\left( {L - x} \right)}}} \right\} + O\left( {{\rm{P}}{{\rm{e}}^{ - 1}},\frac{{{\alpha _L}}}{{{\rm{Pe}}L}}} \right).
\end{equation}
Here, ${U_l} = 1 + \frac{{{\alpha _L}}}{2}$ and ${\lambda _l} = {\rm{Pe}}\sqrt {\frac{{U_l^2}}{d} + \frac{{\left( {d - 1} \right)}}{{{\rm{Pe}}}}} $ are the speed and the inverse boundary layer thickness at the left wall. The corresponding quantities at the right wall are ${U_r} = 1 - \frac{{{\alpha _L}}}{2}$ and ${\lambda _r} = {\rm{Pe}}\sqrt {\frac{{U_r^2}}{d} + \frac{{\left( {d - 1} \right)}}{{{\rm{Pe}}}}} $.

On the other hand, in the presence of a field that orients the particles and for field strengths $\chi_R \ll \lambda \sim {\rm{Pe}}$, the asymptotic solution expanded in terms of $\lambda^{-1} \sim {\rm{Pe}}^{-1}$ is
\begin{equation}
n = {n^{\left( 0 \right)}} + \frac{1}{\lambda }{n^{\left( 1 \right)}} + O\left( {{\rm{P}}{{\rm{e}}^{ - 2}},\frac{{{\chi _R}}}{{{\rm{P}}{{\rm{e}}^2}}}} \right),\,\,{m_x} = m_x^{\left( 0 \right)} + \frac{1}{\lambda }m_x^{\left( 1 \right)} + O\left( {{\rm{P}}{{\rm{e}}^{ - 2}},\frac{{{\chi _R}}}{{{\rm{P}}{{\rm{e}}^2}}}} \right).
\end{equation}
Here, the leading order solution is
\begin{equation}
\frac{{{n^{\left( 0 \right)}}}}{{n_c^{\left( 0 \right)}}} = {e^{{\chi _R}\left( {d - 1} \right)\left( {x - L/2} \right)}} + \frac{{{\rm{Pe}}}}{{d\left( {d - 1} \right)}}\left\{ {{e^{ - {\chi _R}\left( {d - 1} \right)L/2}}{e^{ - \lambda x}} + {e^{{\chi _R}\left( {d - 1} \right)L/2}}{e^{ - \lambda \left( {L - x} \right)}}} \right\},
\end{equation}
\begin{equation}
\frac{{m_x^{\left( 0 \right)}}}{{n_c^{\left( 0 \right)}}} = \frac{\lambda }{{d\left( {d - 1} \right)}}\left\{ { - {e^{ - {\chi _R}\left( {d - 1} \right)L/2}}{e^{ - \lambda x}} + {e^{{\chi _R}\left( {d - 1} \right)L/2}}{e^{ - \lambda \left( {L - x} \right)}}} \right\}
\end{equation}
while the first order solution is
\begin{multline}
{n^{\left( 1 \right)}} = {e^{ - {\chi _R}\left( {d - 1} \right)L/2}}{e^{ - \lambda x}}\left\{ \begin{split}
 -&\, \frac{{{\chi _R}{\rm{P}}{{\rm{e}}^3}n_c^{\left( 0 \right)}}}{{2\lambda {d^2}}}x\\
 +&\, \frac{{{\mathop{\rm Pe}\nolimits} }}{{d\left( {d - 1} \right)}}\left( {n_c^{\left( 1 \right)} + \frac{{{\chi _R}\lambda Ld{{\left( {d - 1} \right)}^2}n_c^{\left( 0 \right)}}}{{2{\rm{Pe}}}} - \frac{{{\chi _R}{\rm{Pe}}n_c^{\left( 0 \right)}}}{d}} \right)
\end{split} \right\}\\
 + {e^{{\chi _R}\left( {d - 1} \right)\left( {x - \frac{L}{2}} \right)}}\left( {n_c^{\left( 1 \right)} - \frac{{{\chi _R}\lambda d{{\left( {d - 1} \right)}^2}n_c^{\left( 0 \right)}}}{{{\rm{Pe}}}}\left( {x - \frac{L}{2}} \right)} \right)\\
 + {e^{{\chi _R}\left( {d - 1} \right)L/2}}{e^{ - \lambda \left( {L - x} \right)}}\left\{ \begin{split}
&\frac{{{\chi _R}{\rm{P}}{{\rm{e}}^3}n_c^{\left( 0 \right)}}}{{2\lambda {d^2}}}\left( {L - x} \right)\\
\,\,\,\, +&\, \frac{{{\rm{Pe}}}}{{d\left( {d - 1} \right)}}\left( {n_c^{\left( 1 \right)} - \frac{{{\chi _R}\lambda Ld{{\left( {d - 1} \right)}^2}n_c^{\left( 0 \right)}}}{{2{\rm{Pe}}}} + \frac{{{\chi _R}{\rm{Pe}}n_c^{\left( 0 \right)}}}{d}} \right)
\end{split} \right\},
\end{multline}
\begin{multline}
m_x^{\left( 1 \right)} = {e^{ - {\chi _R}\left( {d - 1} \right)L/2}}{e^{ - \lambda x}}\left\{ \begin{split}
&\frac{{{\chi _R}{\rm{P}}{{\rm{e}}^2}n_c^{\left( 0 \right)}}}{{2{d^2}\lambda }}\left( {\lambda x - 1} \right)\\
 -&\, \frac{\lambda }{{d\left( {d - 1} \right)}}\left( {n_c^{\left( 1 \right)} + \frac{{{\chi _R}\lambda Ld{{\left( {d - 1} \right)}^2}n_c^{\left( 0 \right)}}}{{2{\rm{Pe}}}} - \frac{{{\chi _R}{\rm{Pe}}n_c^{\left( 0 \right)}}}{d}} \right)
\end{split} \right\}\\
 + \frac{{{\chi _R}\lambda \left( {d - 1} \right)n_c^{\left( 0 \right)}}}{{{\rm{Pe}}}}{e^{{\chi _R}\left( {d - 1} \right)\left( {x - \frac{L}{2}} \right)}}\\
 + {e^{{\chi _R}\left( {d - 1} \right)L/2}}{e^{ - \lambda \left( {L - x} \right)}}\left\{ \begin{split}
&\frac{{{\chi _R}{\rm{P}}{{\rm{e}}^2}n_c^{\left( 0 \right)}}}{{2{d^2}\lambda }}\left( {\lambda \left( {L - x} \right) - 1} \right)\\
 +&\, \frac{\lambda }{{d\left( {d - 1} \right)}}\left( {n_c^{\left( 1 \right)} - \frac{{{\chi _R}\lambda Ld{{\left( {d - 1} \right)}^2}n_c^{\left( 0 \right)}}}{{2{\rm{Pe}}}} + \frac{{{\chi _R}{\rm{Pe}}n_c^{\left( 0 \right)}}}{d}} \right)
\end{split} \right\}.
\end{multline}
The concentration at the center of the confinement at leading and first order, $n_c^{\left( 0 \right)}$, $n_c^{\left( 1 \right)}$, respectively, can be found from the constraints $\frac{1}{L}\int_0^L {{n^{\left( 0 \right)}}dx}  = \left\langle n \right\rangle $ and $\int_0^L {{n^{\left( 1 \right)}}dx}  = 0$. In addition to the constraint $\chi_R \ll {\rm{Pe}}$, we need also $\chi_R \le O\left( 1 \right)$ for this theory to hold because otherwise the nematic order becomes large enough to invalidate the zero nematic order closure based on which this theory is built.

\begin{figure}[tbh!]
\subfloat{\includegraphics[scale = 0.61]{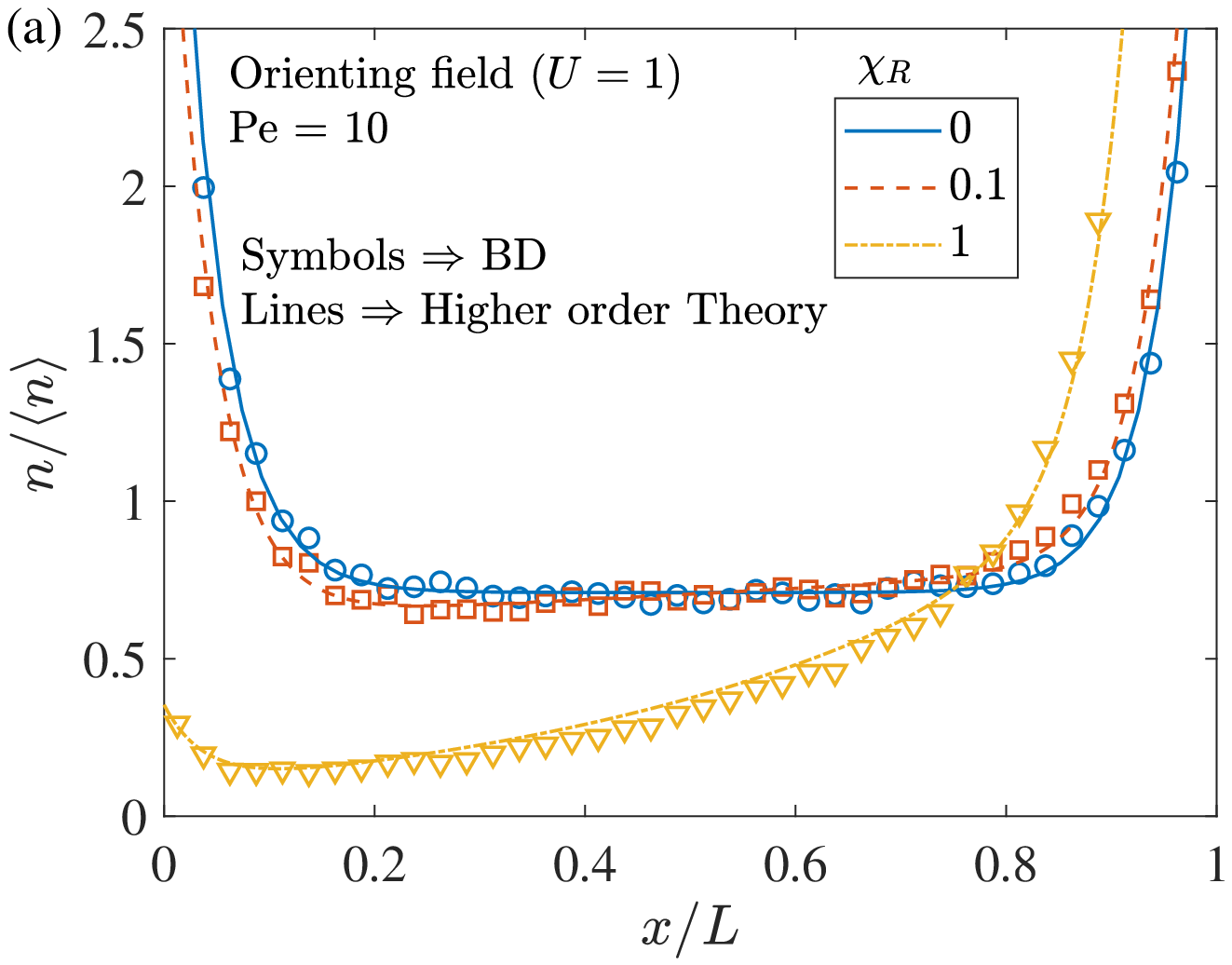}\label{fig:fig1a_sm}}
\subfloat{\includegraphics[scale = 0.61]{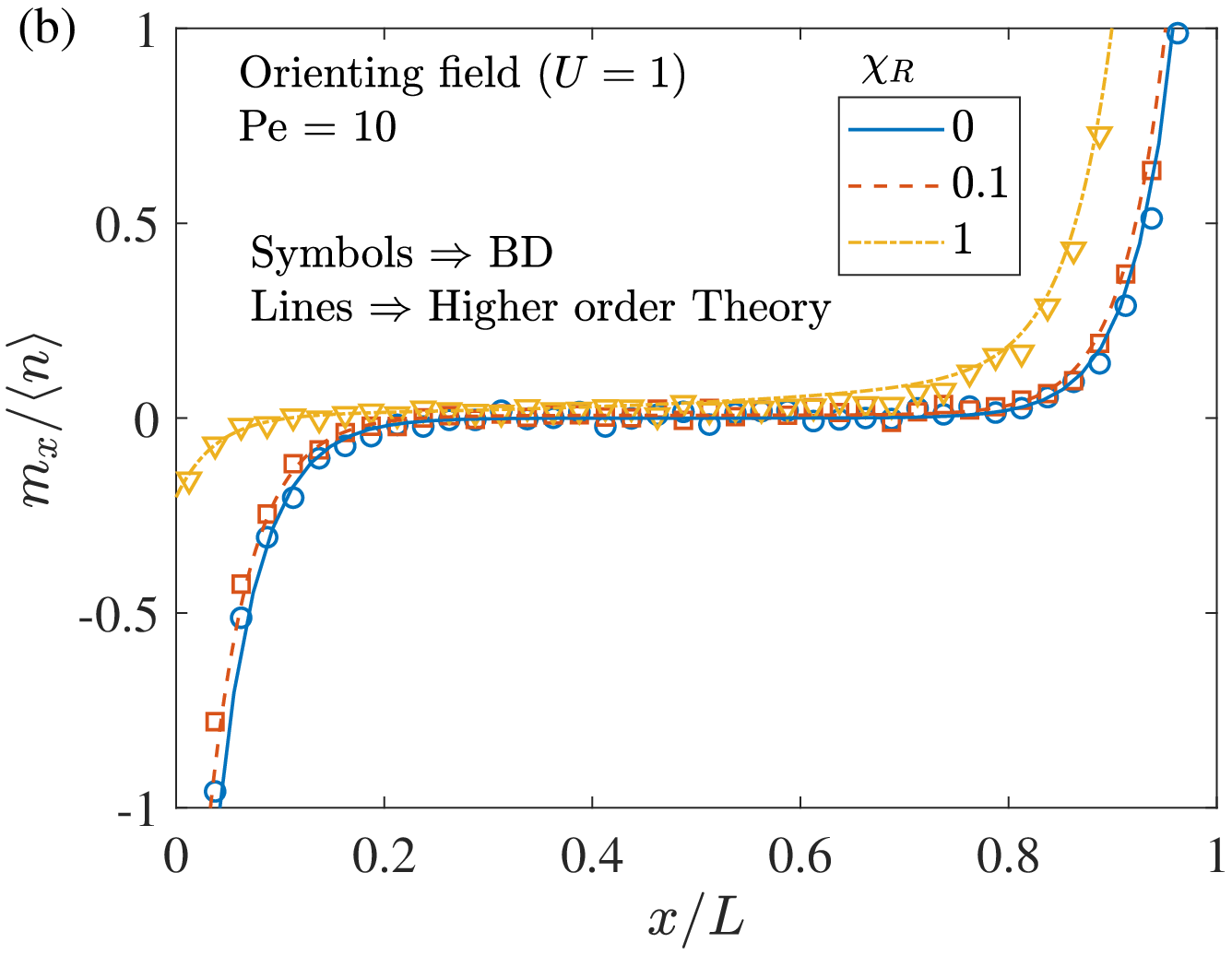}\label{fig:fig1b_sm}}
\caption{\label{fig:fig1_sm}The number density (a), and the polar order (b) associated with the active matter subjected to the orienting field. The symbols denote the BD simulation results while the lines represent the higher order theory. The confinement region $L$ is 10 times larger than the microscopic length $h$.}
\end{figure}

In the main text, we only considered the leading order solution to develop a simple theory. But the accuracy of this theory and hence the match with Brownian Dynamics (BD) simulations can be improved by considering the next order solution. For instance, the number density and polar order reported in Fig.~\ref{fig:fig4} in the main text become those shown in Fig.~\ref{fig:fig1_sm} here, upon inclusion of the next order solution; the improvement in matching with the BD simulations is apparent.

If the orienting field is not constant and varies spatially, ${\bf{H}} = H\left(x\right) \hat{\bf{H}}$, then the asymptotic solution for field strengths $\chi_R \ll \lambda \sim \rm{Pe}$ is
\begin{equation}
\frac{n}{n_c} = e^{-\chi_R \left(d - 1\right) G\left(L/2\right)}\left\{ \begin{aligned} &\,\,\,\,\,\,\,\,\,\,\,\,\,\,\,\,\,\,\,\,\,\,\,\,\,\,\,\,\,\,\,\,\,\,\,\,\,e^{\chi_R \left(d - 1\right) G\left(x\right)}  \\ &+ \frac{{\rm{Pe}}}{d\left( d - 1 \right)} \left( e^{\chi_R \left( d - 1 \right) G\left( 0 \right)}  e^{-\lambda x} + e^{\chi_R \left( d - 1 \right) G\left( L \right)} e^{\lambda \left( x - L \right)} \right)\end{aligned} \right\} + O\left( {\rm{Pe}}^{-1}, \frac{\chi_R}{{\rm{Pe}}} \right),
\end{equation}
\begin{equation}
\frac{m_x}{n_c} = \frac{\lambda e^{-\chi_R \left(d - 1\right) G\left(L/2\right)}}{d \left( d - 1 \right)} \left\{ - e^{\chi_R \left( d - 1 \right) G\left( 0 \right)}  e^{-\lambda x} + e^{\chi_R \left( d - 1 \right) G\left( L \right)} e^{\lambda \left( x - L \right)} \right\} + O\left( {\rm{Pe}}^{-1}, \frac{\chi_R}{{\rm{Pe}}} \right).
\end{equation} 
Here, $G\left(x\right) = \int{H\left(x\right) dx}$ and the concentration at the center of the confinement $n_c$ can again be found from the constraint $\frac{1}{L}\int_0^L {n dx}  = \left\langle n \right\rangle $.

\section{\label{app:BD}Brownian Dynamics simulations}
The Brownian Dynamics simulations reported in the main text are carried out by numerically integrating over-damped Langevin equations in time \citep{Graham2018}
\begin{equation}
{\bf{0}} =  - \zeta {\bf{\dot x}} + {{\bf{F}}^{swim}} + {{\bf{F}}^B},
\end{equation}
\begin{equation}
{\bf{0}} =  - {\zeta _R}{\bf{\Omega }} + {{\bf{L}}^{ext}} + {{\bf{L}}^R},
\end{equation}
where the particle orientation ${\bf{q}}$ follows $\frac{{d{\bf{q}}}}{{dt}} = {\bf{\Omega }} \times {\bf{q}}$. Here $\zeta$, $\zeta_R$, are the the translational and rotational resistances. The swim force ${{\bf{F}}^{swim}} = \zeta U\left( {\bf{x}} \right){\bf{q}}$ and the torque exerted by the orienting field ${{\bf{L}}^{ext}} = {\zeta _R}{\Omega _c}\left( {{\bf{q}} \times {\bf{H}}} \right)$. The fluctuating force ${{\bf{F}}^B}$ and the torque ${{\bf{L}}^R}$ follow the usual white noise statistics: $\overline {{{\bf{F}}^B}\left( t \right)}  = \bm{0}$, $\overline {{{\bf{F}}^B}\left( 0 \right){{\bf{F}}^B}\left( t \right)}  = 2{k_B}T\zeta \delta \left( t \right){\bf{I}}$, $\overline {{{\bf{L}}^R}\left( t \right)}  = \bm{0}$, $\overline {{{\bf{L}}^R}\left( 0 \right){{\bf{L}}^R}\left( t \right)}  = 2\zeta _R^2\delta \left( t \right){\bf{I}}/{\tau _R}$, where the overbar denotes an ensemble average.

The numerical integration of the Langevin equations is carried out using the Euler-Maruyama scheme with the time-step $\Delta t = {10^{ - 4}}{\tau _R}$ \citep{Kloeden1992}. The simulations are run for $10^5$ particles until the time $t = 100{\tau _R}$. The penetration of particles into the wall is avoided by using the potential-free algorithm \citep{Foss2000}. The wall separation $L$ already includes the particle size, and thus the algorithm simplifies to setting the particle position $x$ to $0$ or $L$, respectively, if $x < 0$ or $> L$.

\section{\label{app:nematic}Nematic order}
\begin{figure}[tbh!]
\subfloat{\includegraphics[scale = 0.40]{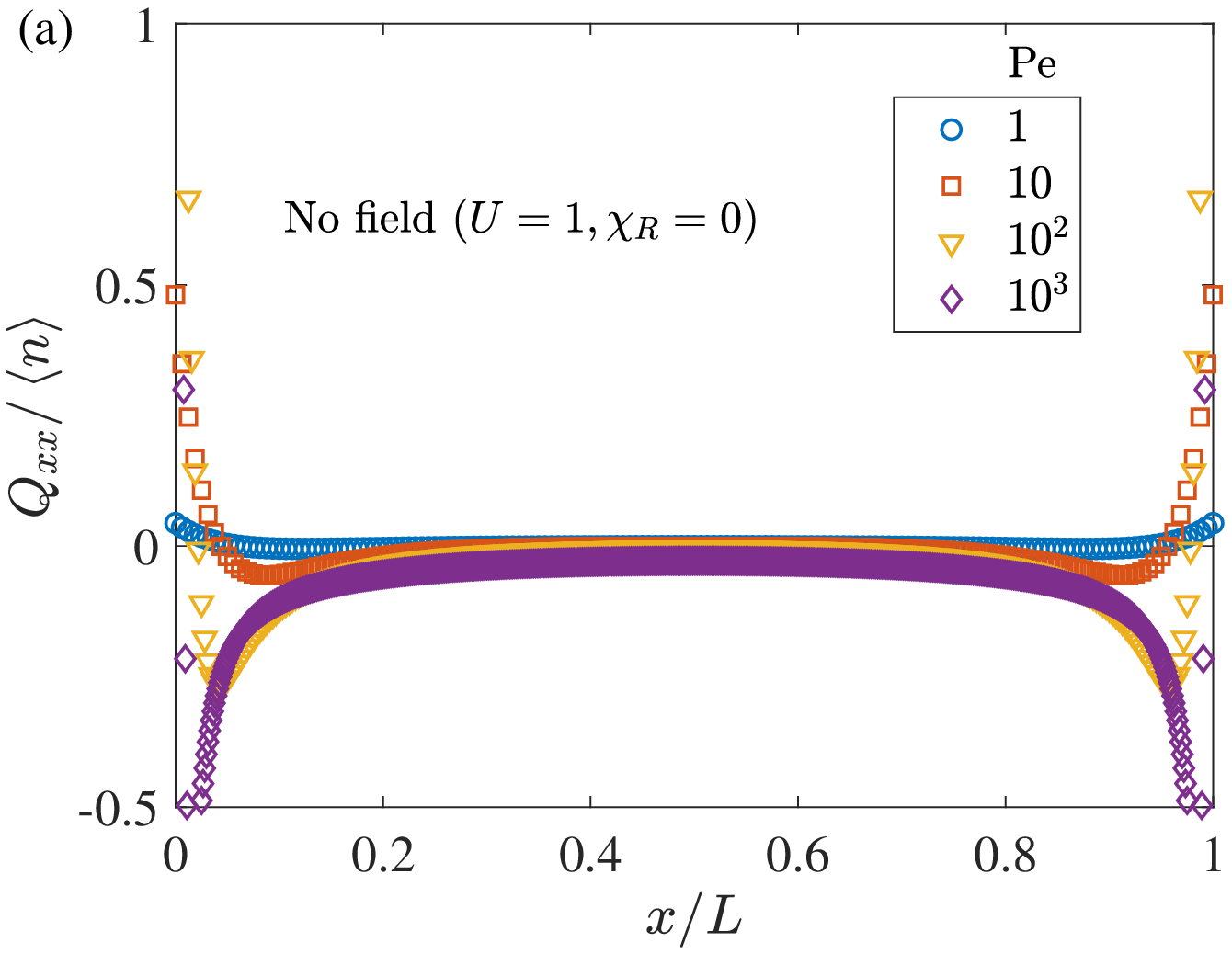}\label{fig:fig2a_sm}}
\subfloat{\includegraphics[scale = 0.40]{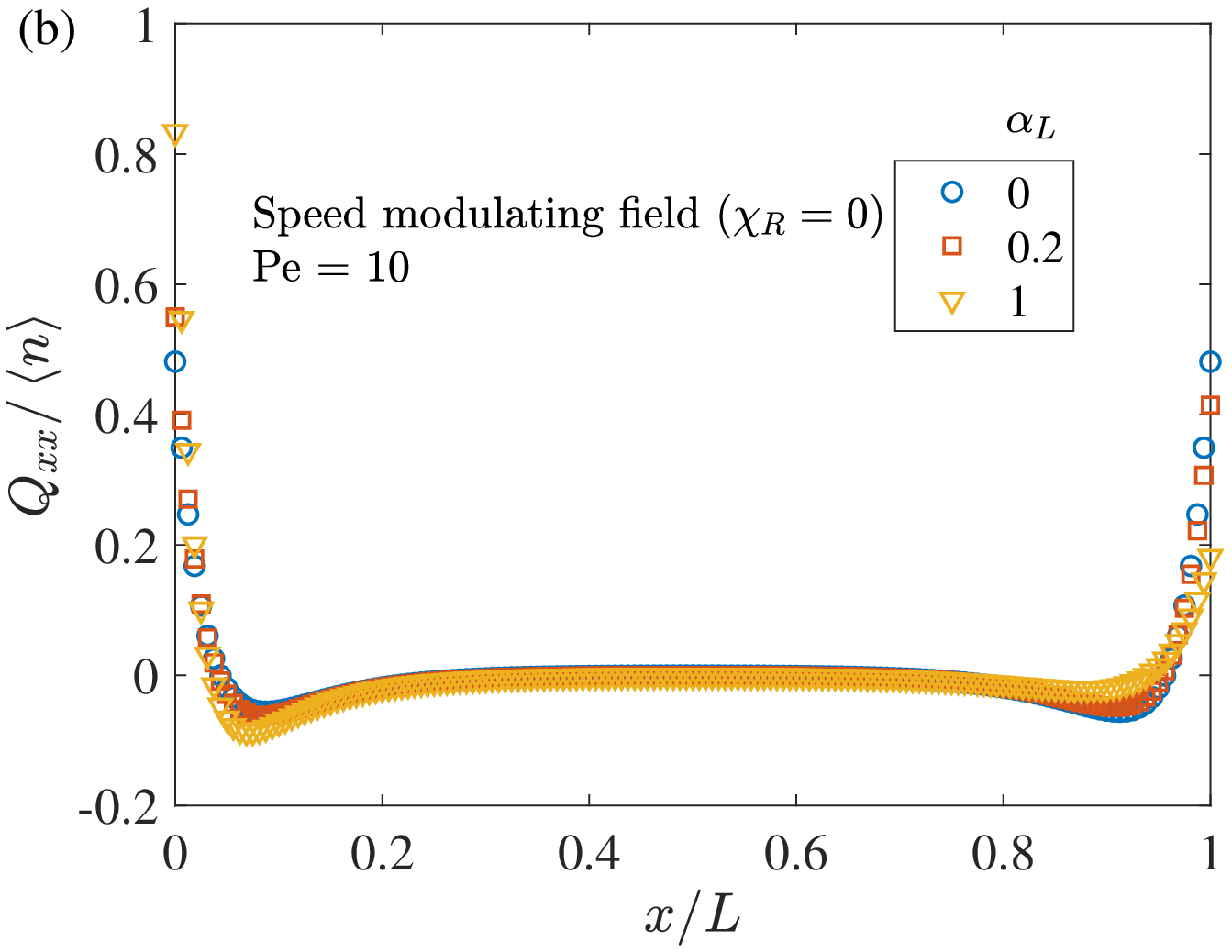}\label{fig:fig2b_sm}}
\subfloat{\includegraphics[scale = 0.40]{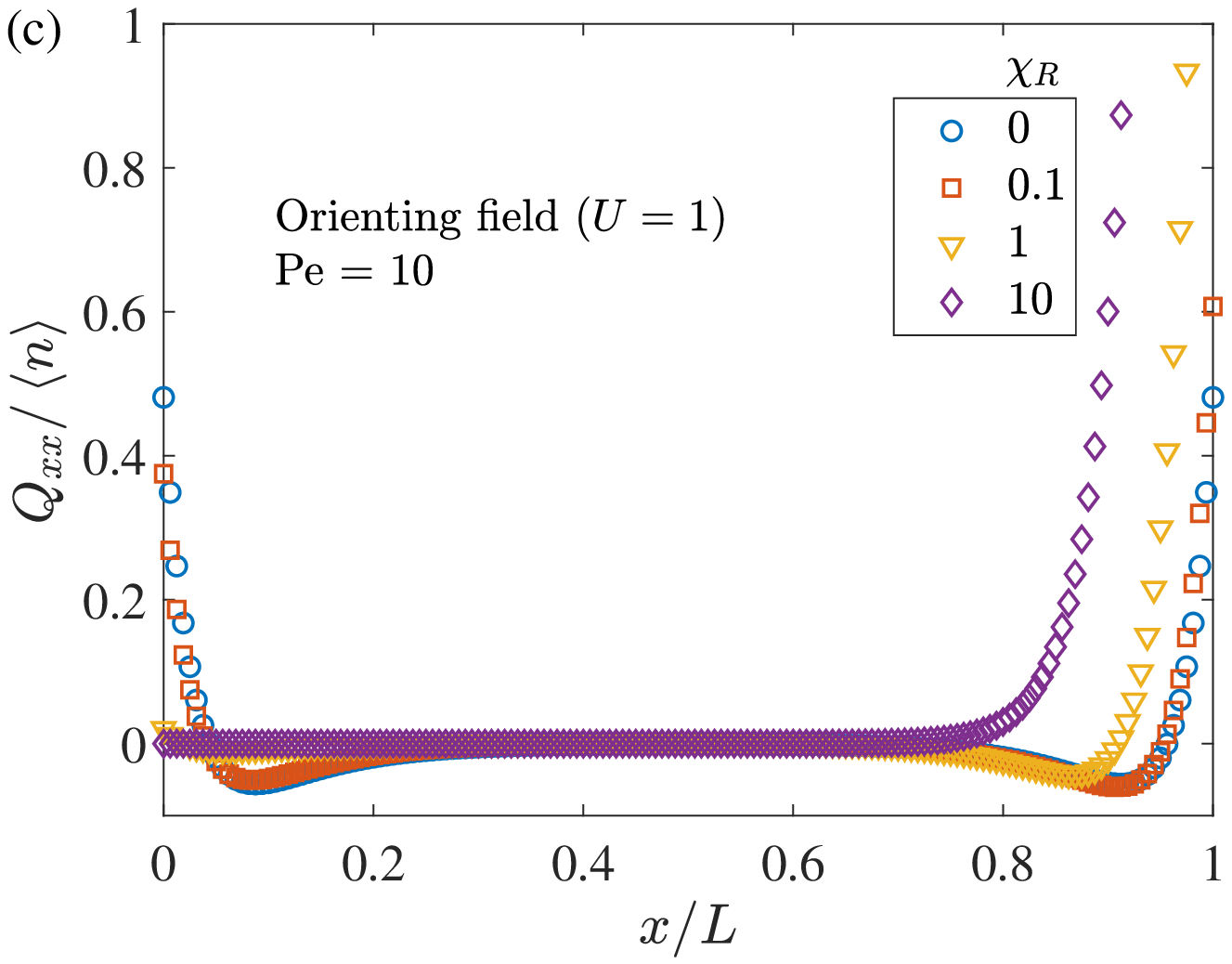}\label{fig:fig2c_sm}}
\caption{\label{fig:fig2_sm}The nematic order associated with the active matter without any field (a), or subjected to a speed modulating field (b) or an orienting field (c). The confinement region $L$ is chosen as $10$ times larger than the microscopic length $h$.}
\end{figure}

We also solved the Smoluchowski equation numerically using in-house FEM code. The numerical solution yields the probability density, from which its moments were evaluated. The first two moments, the number density and the polar order, computed are consistent with theory and BD simulations. Typical values of the next moment, nematic order, at various Pe, $\chi_R$ and $\alpha_L$ are shown in Fig.~\ref{fig:fig2_sm}. Nematic order is small and hence we assume it is safe to neglect for ${\rm{Pe}} < 10^3 $ and field strengths $\alpha_L < 1, \chi_R < 1$.

\bibliography{references}

\end{document}